\documentclass[printer]{aa} % for a referee version
\usepackage{graphicx}
\usepackage{txfonts}
\usepackage[switch]{lineno}
\usepackage[shortlabels]{enumitem}

\begin{document} 

%\linenumbers

%

\title{Improving pulsar polarization and timing measurements with the Nan\c{c}ay Radio Telescope}

\author{
L. Guillemot\inst{1,2}
\and
I. Cognard\inst{1,2}
\and
W. van Straten\inst{3,4}
\and
G. Theureau\inst{1,2,5}
\and
E. G\'erard\inst{6}
}

\institute{
Laboratoire de Physique et Chimie de l'Environnement et de l'Espace, Universit\'e d'Orl\'eans / CNRS, 45071 Orl\'eans Cedex 02, France \\
\email{lucas.guillemot@cnrs-orleans.fr}
\and
Observatoire Radioastronomique de Nan\c{c}ay, Observatoire de Paris, Universit\'e PSL, Université d'Orl\'eans, CNRS, 18330 Nan\c{c}ay, France
\and
Institute for Radio Astronomy \& Space Research, Auckland University of Technology, Private Bag 92006, Auckland 1142, New Zealand
\and
Manly Astrophysics, 15/41-42 East Esplanade, Manly, NSW 2095, Australia
\and
LUTH, Observatoire de Paris, Universit\'e PSL, Universit\'e Paris Cit\'e, CNRS, 92195 Meudon, France
\and
GEPI, Observatoire de Paris, Universit\'e PSL, Universit\'e Paris Cit\'e, CNRS, 92195 Meudon, France
}

\date{Received 26 May 2023 / Accepted 2 August 2023}

\authorrunning{L. Guillemot et al.} 

\abstract
{Accurate polarimetric calibration of the radio pulse profiles from pulsars is crucial for studying their radiation properties at these wavelengths. Additionally, inaccurate calibration can distort recorded pulse profiles, introducing noise in time of arrival (TOA) data and thus degrading pulsar timing analyses. One method for determining the full polarimetric response of a given telescope is to conduct observations of bright polarized pulsars over wide ranges of parallactic angles, to sample different orientations of their polarization angle and in turn determine the cross-couplings between polarization feeds.}
{The Nan\c{c}ay decimetric Radio Telescope (NRT) is a 94m equivalent meridian telescope, capable of tracking a given pulsar for approximately one hour around transit. The NRT therefore cannot sample wide ranges of parallactic angles when observing a given pulsar, so until late 2019 the polarimetric calibration of 1.4~GHz pulsar observations with the NRT was rudimentary. We therefore aimed to develop a method for improving the calibration of NRT observations, overcoming the above-mentioned limitation. Ultimately, our goal was to improve the quality of NRT pulsar timing, with better calibrated pulsar pulse profiles.}
{In November 2019, we began conducting regular observations of the bright and highly linearly polarized pulsar PSR~J0742$-$2822, in a special observing mode in which the feed horn rotates by $\sim 180^\circ$ over the course of the one hour observation, mimicking wide parallactic angle variations and in principle enabling us to determine the polarimetric response of the NRT at 1.4~GHz. In addition, we assessed the quality of the NRT timing of a selection of millisecond pulsars (MSPs), namely, J1730$-$2304, J1744$-$1134, and J1857+0953, with conventional TOAs extracted from total intensity pulse profiles, and TOAs extracted with the Matrix Template Matching (MTM) technique, designed to compensate for putative polarimetric calibration errors.}
{From the analysis of the rotating horn observations of PSR~J0742$-$2822 we could determine the cross-couplings between the polarization feeds and also constrain the Stokes parameters of the noise diode signal, which prior to this work was erroneously assumed to be ideal and was used as the only reference source for the calibration of pulsar observations. The improved polarimetric response of the NRT as determined from these observations was applied to observations of a selection of MSPs with published polarimetric properties. We find that the new polarimetric profiles and polarization position angles are consistent with previous findings, unlike NRT polarimetric results obtained with the previously used method of calibration. The analysis of the timing data shows that the new calibration method improves the quality of the timing, and the MTM method proves very effective at reducing noise from imperfect calibration. For pulsars with sufficient degrees of polarization, the MTM method appears to be the preferred method of extracting TOAs from NRT observations.}
{}


\maketitle

%%%

\section{Introduction}
\label{sec:intro}

Pulsars are rapidly rotating, highly magnetized neutron stars that emit beams of electromagnetic radiation. These beams are swept across the sky as they rotate, so that the emission appears to be ``pulsed'' to distant observers whose lines of sight are crossed by the beams. Pulsars can be used as tools to probe a variety of astrophysical topics, via the ``pulsar timing'' technique. Pulsar timing consists of measuring pulse times of arrival (TOAs), which correspond to epochs at which a fiducial phase of the pulsar's periodic signal is detected at a telescope. The TOAs of a given pulsar encode much information about its spin properties, its orbital properties if the pulsar belongs to a multiple system, information about the intervening interstellar medium, and other characteristics of the pulsar and its environment \citep[see, for instance,][for reviews of pulsar timing and its applications]{Taylor1992,Stairs2003,Handbook}. These properties can be determined by minimizing the differences between the TOAs and the predictions from a model for the pulsar encoding the above-listed information, the so-called ``timing residuals''. For some pulsars with high rotational stability, timing residuals with standard deviations on the order of 100~ns can be achieved \citep[see e.g.,][]{Liu2020}.

The Nan\c{c}ay Radio Telescope (NRT) is a meridian transit-type telescope of the Kraus/Ohio State design \citep[see e.g.,][]{Kraus1960,Kraus1966}. It consists of a tiltable flat primary mirror (with dimensions 200 $\times$ 40m), a fixed spherical secondary mirror (300 $\times$ 35m) separated from the primary mirror by 460m, and a receiver carriage that follows the focal point during observations by moving on a 80m train track. The NRT is equivalent to a parabolic dish with a diameter of about 94m, and can track objects with declinations above $\sim -39^\circ$ for approximately one hour around transit. It is equipped with two cryogenically cooled receivers that respectively cover the $\left[1.1; 1.8\right]$ GHz and $\left[1.7; 3.5\right]$ GHz frequency ranges, allowing for measurement of the four Stokes parameters over those ranges. Since August 2011, pulsar observations with the NRT have been conducted using the NUPPI backend, a version of the Green Bank Ultimate Pulsar Processing Instrument used at the Green Bank Telescope \citep{DuPlain2008}, and designed for the NRT. The NUPPI instrument \citep[for further details, see e.g.,][]{Desvignes2011,Cognard2013} uses Graphics Processing Units (GPUs) to coherently dedisperse and fold the dual linear polarization signals from the receiver in real time, over a total bandwidth of 512~MHz. The 512~MHz bandwidth is split into 128 channels of 4~MHz each, and in practice the bulk of pulsar observations with the NUPPI backend are conducted at a central frequency of 1484~MHz, to exploit as much of the frequency interval covered by the 1.1--1.8\,GHz receiver as possible. A smaller fraction of the observations is conducted with the high-frequency receiver at a central frequency of 2539~MHz, or at other central frequencies, depending on the pulsar or science objective. Brief descriptions of the instruments used for conducting pulsar timing observations with the NRT prior to NUPPI can be found in \citet{Desvignes2016}. 

Pulsar timing data recorded with the NUPPI backend feature prominently in several recent high-precision pulsar timing studies. Examples include the timing of PSR~J0737$-$3039A in the so-called ``double pulsar'' system, enabling highly precise tests of General Relativity (GR) in the strong-field regime \citep{Kramer2021}, tests of the strong equivalence principle of GR via the timing of PSR~J0337+1715 in a stellar triple system \citep{Voisin2020}, the measurement of the mass of PSR~J2222$-$0137 in a massive neutron star -- white dwarf binary system \citep{Cognard2017,Guo2021}, the timing of the eclipsing black widow pulsar PSR~J2055+3829 discovered at Nan\c{c}ay \citep{Guillemot2019}, or searches for low-frequency gravitational waves using Pulsar Timing Arrays \citep[PTAs; see e.g.,][]{Chen2021}. All of these studies analyzed TOAs that had been extracted from NUPPI data by cross-correlating a standard pulse profile for the pulsar of interest with individual observations, partially averaged in frequency and/or time. A comprehensive review of the TOA extraction process and of sources of measurement uncertainties can be found in e.g., \citet{Verbiest2018}, and a recent comparison of TOA creation practices using data from the NRT and other telescopes can be found in \citet{Wang2022}. One crucial assumption made when extracting TOAs from timing data by cross-correlating a standard (or template) pulse profile with observations of a given pulsar is that the observations resemble the standard profile; that is, they are scaled versions of the template profile on top of varying white noise levels. Distortions of the observed pulse profile caused by time-varying instrumental artifacts can cause individual observations to differ from the template profile, affecting the accuracy of the TOAs and introducing systematic timing errors in the TOA dataset. As a consequence, the individual pulsar timing observations need to be carefully calibrated in polarization. Moreover, accurate polarization properties of a given pulsar can provide a useful insight into the physical processes at work in its magnetosphere \citep[see e.g.,][]{Philippov2022}.

As will be presented in further detail in Sect.~\ref{sec:data}, until recently the procedure followed for calibrating pulsar observations with the NRT was limited. With seemingly correct pulsar timing results with the NRT (as e.g., determined in the context of multi-telescope pulsar timing studies), and a relative paucity of pulsar polarization studies with this telescope \citep[see for instance][for examples of NRT pulsar polarization measurements]{Theureau2011,Dyks2016,Desvignes2022}, the validity of the former polarimetric calibration procedure had not been studied in detail prior to this work. However, detailed comparisons of NRT polarization profiles with reference results obtained at other telescopes, and significant modifications in the observed polarization properties of pulsars concurrent with instrumental changes in 2019, prompted us to revisit the procedure for calibrating pulsar observations with the NRT, and the procedure for extracting TOAs from NRT data. In Sect.~\ref{sec:data} we present efforts to improve the polarimetric calibration procedure. In Sect.~\ref{sec:timing} we present the 1.4~GHz NUPPI data archive, and assess the impact of the improved calibration procedure on the quality of the timing, investigating a novel use of the Matrix Template Matching technique \citep[MTM; see][]{vanStraten2006} with NRT data. This technique uses the timing information in all four Stokes components of the pulsar signal while compensating for residual polarimetric calibration errors when extracting TOAs from observations. Finally, we present a summary of our findings and some prospects from this study in Sect.~\ref{sec:summary}. 

%%%%%%%%%%%%%%%

\section{Observations and polarimetric calibration}
\label{sec:data}

\subsection{First order polarimetric calibration of NUPPI observations}
\label{subsec:calib1}

In August 2011, the NUPPI backend became the primary pulsar instrument in operation at the NRT, replacing the Berkeley-Orl\'eans-Nan\c{c}ay (BON) pulsar backend. As mentioned in the introduction, the majority of NRT pulsar timing observations are conducted at a central frequency of 1.484~GHz with the low-frequency receiver. In this work we focus on these observations, leaving observations with previous backends or with the high-frequency receiver for future work.

\begin{figure}[ht!]
\begin{center}
\includegraphics[width=0.95\columnwidth]{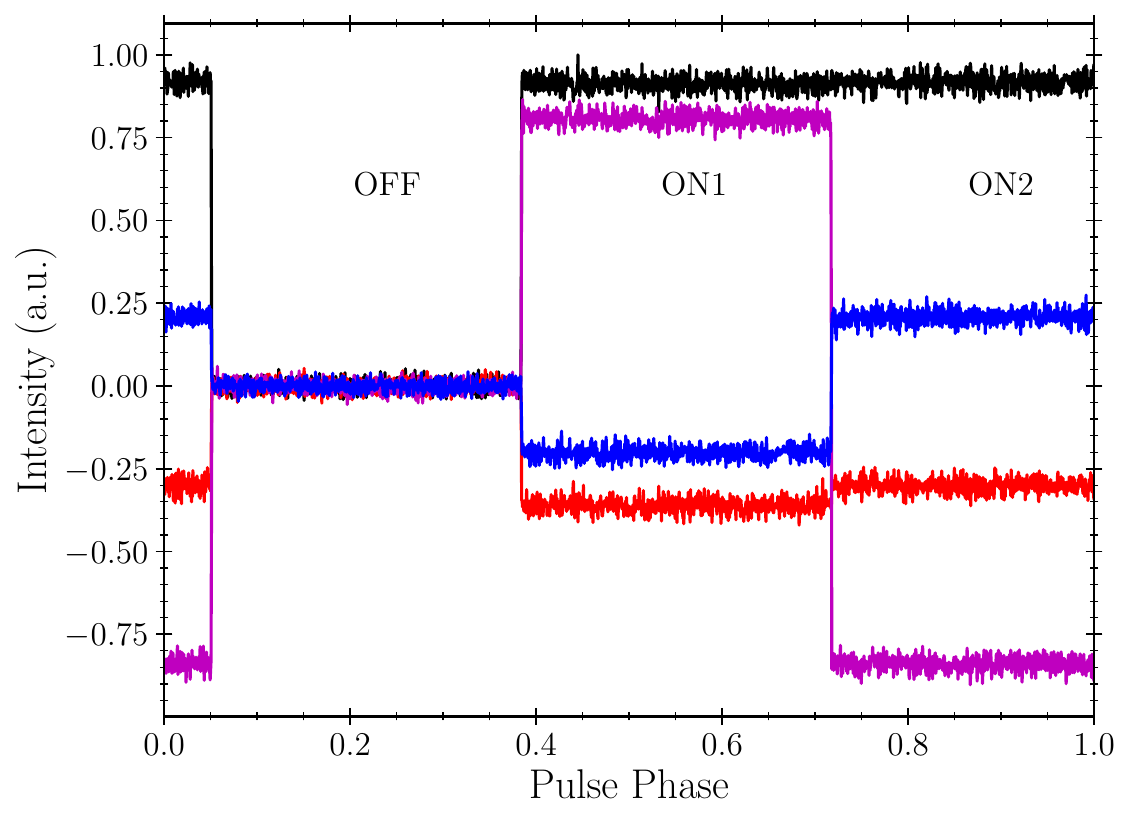}
\caption{Example of a noise diode observation conducted prior to a pulsar observation with the NRT. The black line represents the total intensity, $I$, of the signal as a function of pulse phase (the noise diode signal is periodic, with a frequency of 3.33 Hz), and the red, magenta, and blue lines correspond respectively to the $Q$, $U$, and $V$ Stokes parameters. Note that the $I$, $Q$, $U$, and $V$ Stokes parameters are uncalibrated in this figure. The data were extracted from a 10-s observation of the noise diode with the NUPPI backend, conducted on MJD~59368 prior to an observation of PSR~J0742$-$2822 at the central frequency of 1.484~GHz. The data shown in this figure were taken from the 46th 4-MHz channel recorded by NUPPI, corresponding to the frequency of 1.410~GHz.}
\label{fig:cal_0742}
\end{center}
\end{figure}

At the beginning of each pulsar observation with the NRT, a 3.33~Hz noise diode injects a polarized reference signal into the receiver feed horn for ten seconds. An example of a noise diode observation is displayed in Figure~\ref{fig:cal_0742}. As can be seen from the figure, the noise diode used at the NRT has three different states: an ``OFF'' state, in which the diode is inactive, and two ``ON'' states. In the first ``ON'' state (\textit{i.e.,} the ``ON1'' state), the noise diode theoretically injects a 100\% linearly polarized signal that illuminates both feeds equally, which corresponds to 100\% Stokes $U$ polarization (note that in Figure~\ref{fig:cal_0742} the noise diode observation is not calibrated, so the ``ON1'' state appears to contain little emission along Stokes $U$, and strong circularly polarized emission). Under the latter assumption, the signal from the noise diode can thus be used to perform a first-order calibration of the complex gains of the two polarization feeds, as implemented in \textit{e.g.,} the \textsc{SingleAxis} calibration method of the PSRCHIVE software library \citep{Hotan2004} used for manipulating the pulsar data presented throughout this article. The ``ON2'' state, on the other hand, theoretically consists of a 100\% linearly polarized signal along Stokes $-U$ and is not used by PSRCHIVE in the calibration procedure.

In ten seconds, the noise diode is typically detected in each 4~MHz channel with a signal-to-noise ratio (S/N) of 2000. Throughout this article, S/N values are calculated using Equation~7.1 of \citet{Handbook}, \textit{i.e.,}

\begin{equation}
\mathrm{S/N} = \dfrac{1}{\sigma_\mathrm{Off} \sqrt{W_\mathrm{eq}}} \sum_{i=1}^{N_\mathrm{bins}} \left(p_i - p_\mathrm{Off}\right),
\label{eq:snr}
\end{equation}

\noindent
where $p_i$ denotes the amplitude of the $i^\mathrm{th}$ bin, $\sigma_\mathrm{Off}$ and $p_\mathrm{Off}$ represent the off-pulse standard deviation and mean amplitude, and $W_\mathrm{eq}$ is the equivalent width of a top-hat pulse with the same area and peak height as the profile, expressed in number of bins. With an S/N of 2000, and given the characteristics of the NRT \citep[see Table~6 of][]{Wu2018}, we estimate, using Equation~7.10 of \citet{Handbook}, that the noise diode increases the receiver temperature by $\sim 20$\% of the system temperature. This significant temperature increase enables accurate determination of the calibration parameters.

A comprehensive description of the formalism employed by PSRCHIVE for the calibration of pulsar observations can be found in \citet{vanStraten2004}. In a nutshell, the \textsc{SingleAxis} method of PSRCHIVE assumes that the polarization feeds are perfectly orthogonal and that the reference signal from the noise diode is 100\% linearly polarized and illuminates the two feeds equally and in phase, as should be the case under the assumption made above. Under these hypotheses, the $2 \times 2$ complex-valued Jones matrix $\boldsymbol{J}_\mathrm{SA}$ describing the response of the instrument can be written as: 

\begin{equation}
\boldsymbol{J}_\mathrm{SA} = G\ \boldsymbol{B}_{\boldsymbol{\hat{q}}} \left(\gamma\right)\ \boldsymbol{R}_{\boldsymbol{\hat{q}}} \left(\varphi\right),
\label{eq:jones_singleaxis}
\end{equation}

\noindent
where $G$ is the absolute gain, $\gamma$ denotes the differential gain between the feeds, $\varphi$ is the differential phase, and $\boldsymbol{B}_{\boldsymbol{\hat{q}}}$ and $\boldsymbol{R}_{\boldsymbol{\hat{q}}}$ denote Lorentz boost and rotation transformations \citep{Britton2000}. Under the assumptions made here, the Jones matrix has the following simple form: 

\begin{equation}
\boldsymbol{J}_\mathrm{SA} = G\ \begin{pmatrix}
e^{\gamma + i \varphi} & 0 \\
0 & e^{-\left(\gamma + i \varphi\right)} 
\end{pmatrix}.
\end{equation}

\begin{figure}[ht]
\begin{center}
\includegraphics[width=0.95\columnwidth]{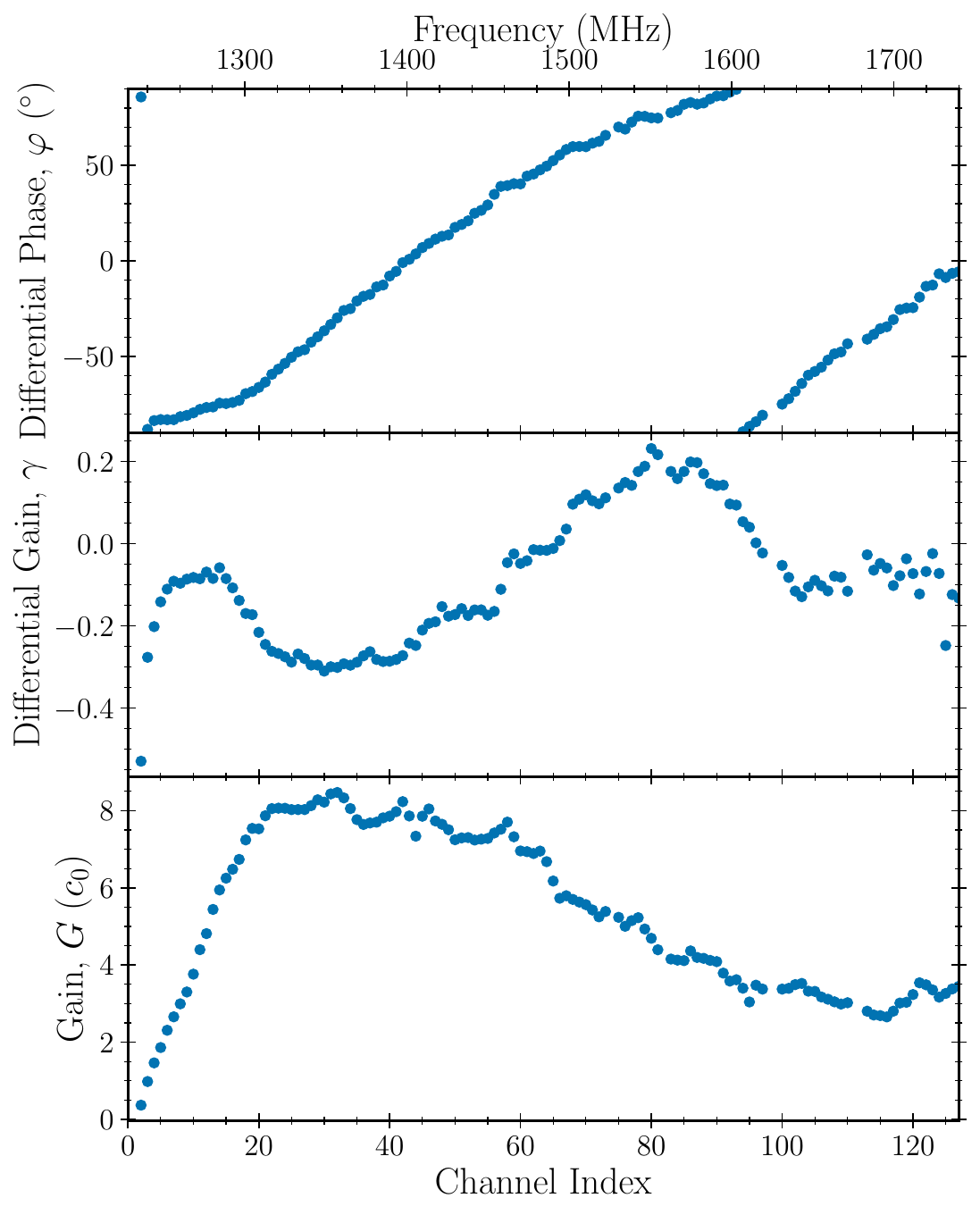}
\caption{First order calibration parameters as determined from the \texttt{SingleAxis} analysis of the noise diode observation shown in Figure~\ref{fig:cal_0742}, as a function of frequency. \textit{(Top)} Differential phase, $\varphi$. \textit{(Middle)} Differential gain, $\gamma$. \textit{(Bottom)} Absolute gain, $G$, expressed in units of the square root of the flux density of the noise diode in the considered frequency channel, $c_0 = \sqrt{C_0}$.}
\label{fig:sol_0742}
\end{center}
\end{figure}

\noindent
The measured coherency matrix, $\boldsymbol{\rho^\prime}_\mathrm{ref}$, of the reference signal from the noise diode is then given by the polarization measurement equation: 

\begin{equation}
\boldsymbol{\rho^\prime}_\mathrm{ref} = \boldsymbol{J}_\mathrm{SA}\ \boldsymbol{\rho}_\mathrm{ref}\ \boldsymbol{J}_\mathrm{SA}^\dagger, 
\label{eq:coherency}
\end{equation}

\noindent
where $\boldsymbol{\rho}_\mathrm{ref}$ is the coherency matrix of the input ideal reference signal, in our case: 

\begin{equation}
\boldsymbol{\rho}_\mathrm{ref} = \dfrac{C_0}{2} \begin{pmatrix}
 1 & 1 \\
 1 & 1 \\
\end{pmatrix}, 
\label{eq:ideal_diode}
\end{equation}

\noindent
with $C_0$ the flux density of the noise diode at the considered frequency. The \textsc{SingleAxis} method uses Equation~\ref{eq:coherency} and measurements of $\boldsymbol{\rho^\prime}_\mathrm{ref}$ at different frequencies to determine $G$, $\gamma$, and $\varphi$ across the bandwidth recorded by NUPPI. Figure~\ref{fig:sol_0742} shows the measured $G$, $\gamma$, and $\varphi$ parameters as a function of frequency, as obtained from the analysis of the noise diode observation plotted in Figure~\ref{fig:cal_0742}.

The \textsc{SingleAxis} Jones matrix characterizing the instrumental response determined from the analysis of the noise diode observation can then be used to perform the polarimetric calibration of the pulsar observation conducted subsequently. The Jones matrix to be used for this transformation is $\boldsymbol{J}_\mathrm{SA} \ \boldsymbol{R}_{\boldsymbol{\hat{v}}} \left( \Phi \right)$, where $\boldsymbol{J}_\mathrm{SA}$ is the Jones matrix describing the instrumental response and $\Phi$ is the time-varying parallactic angle, given in the case of the NRT by:

\begin{equation}
\sin\left(\Phi\right) = \sin\left(\delta\right) \times \sin\left(\mathrm{HA}\right), 
\label{eq:parallactic}
\end{equation}

\noindent
where $\delta$ and HA are the declination and the (time-varying) hour angle of the source. 

\begin{figure*}[ht!]
\begin{center}
\includegraphics[scale=0.45]{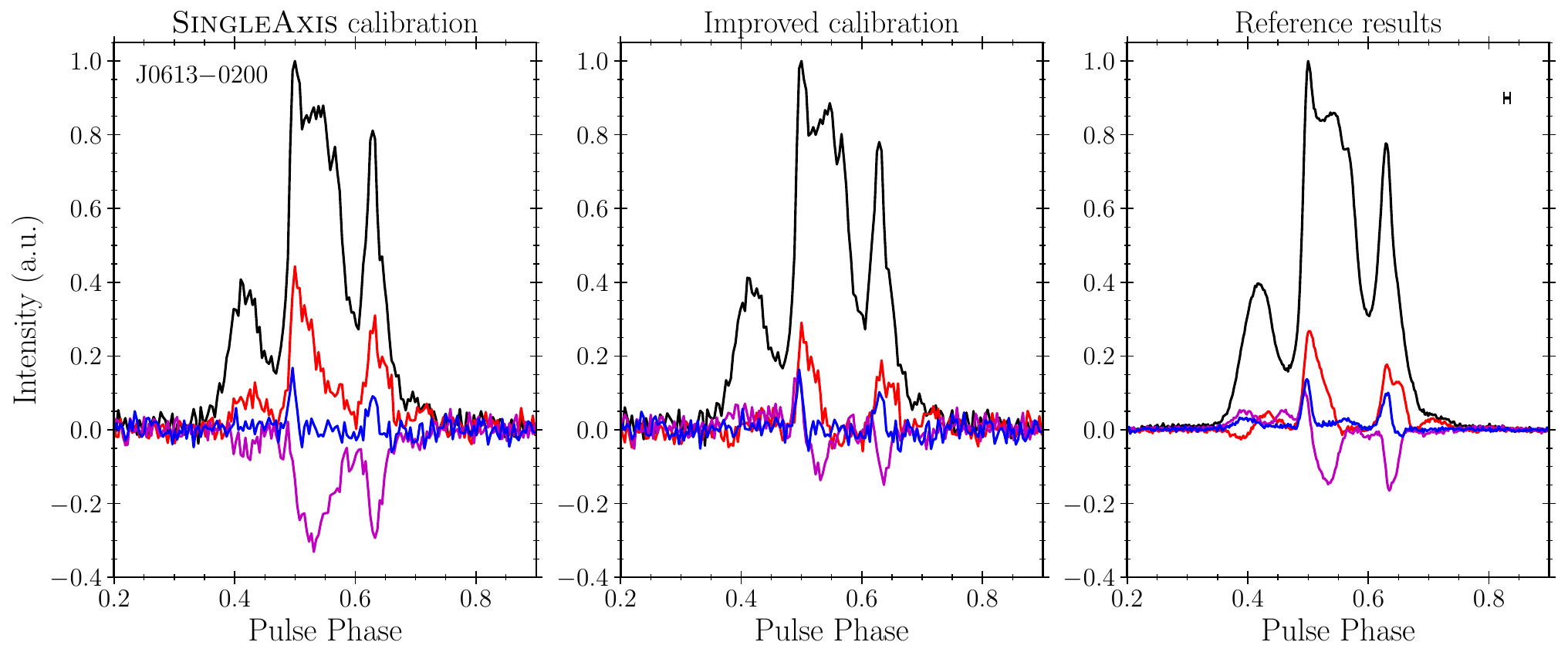}
\includegraphics[scale=0.45]{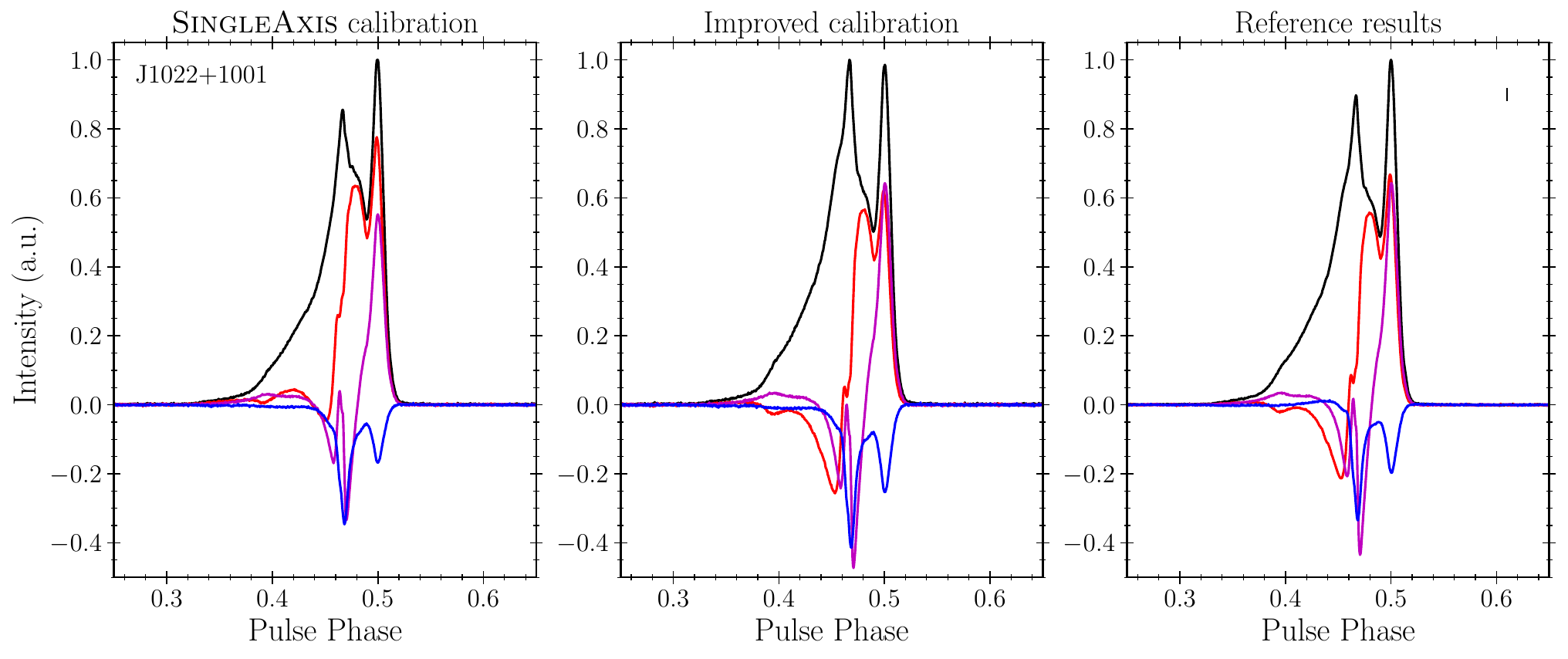}
\includegraphics[scale=0.45]{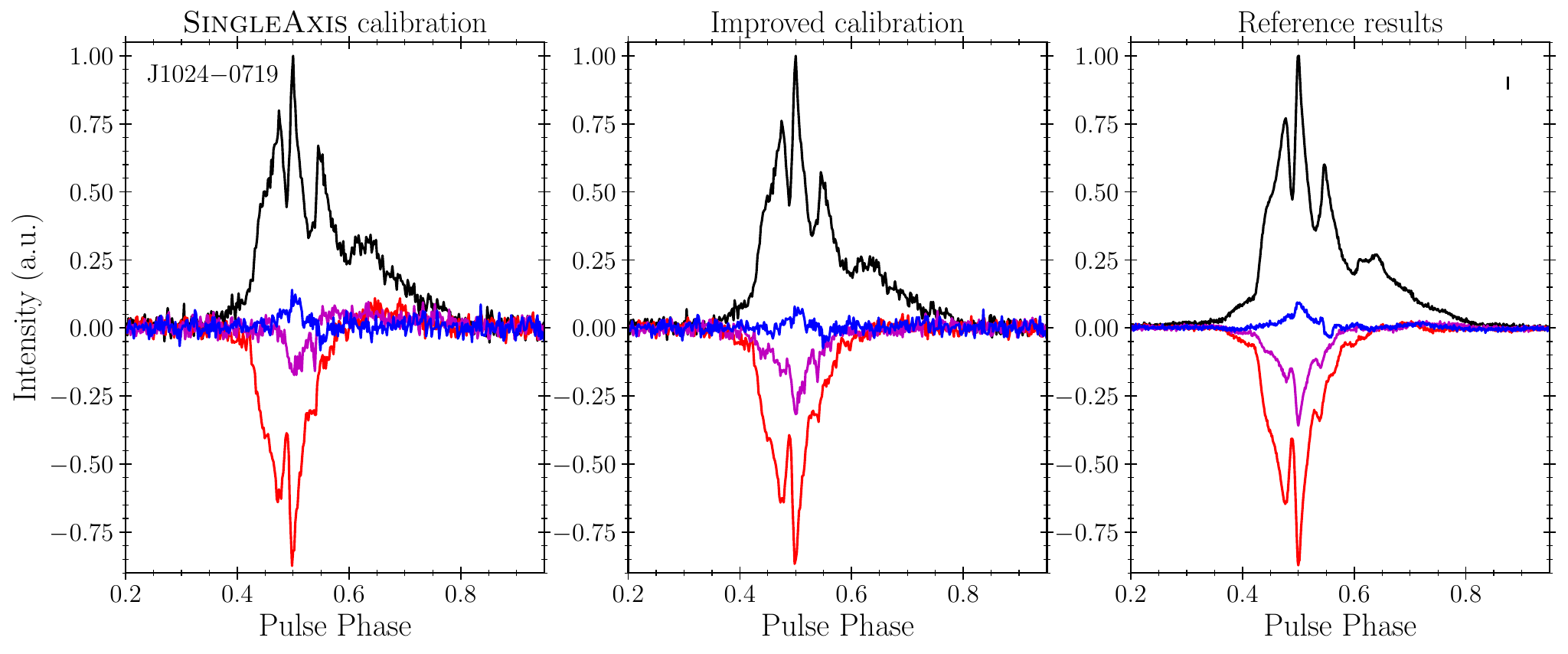}
\caption{Polarimetric profiles for PSRs J0613$-$0200, J1022+1001, and J1024$-$0719. In each panel, the black line represents the total intensity (Stokes parameter $I$), the red and magenta lines correspond to the Stokes parameters $Q$ and $U$ describing the linear polarization, and the blue line is the circular polarization (Stokes parameter $V$). For each pulsar, the left-hand panel shows NUPPI pulse profiles calibrated with the \textsc{SingleAxis} method as described in Section~\ref{subsec:calib1}. The middle panel shows NUPPI polarimetric profiles calibrated with the improved scheme presented in Section~\ref{subsec:calib2}. Reference results from \citet{Dai2015} are displayed in the right-hand panel. In the latter panel, the marker at the top right indicates the dispersion smearing resulting from incoherent dedispersion of the Parkes data. The displayed pulse phase ranges were restricted to intervals with significant emission features, to facilitate comparisons. The profiles were normalized to the maximum value of the total intensity. See Table~\ref{tab:obstable} for details on the NUPPI observations presented in this figure.}
\label{fig:pol_profs_1}
\end{center}
\end{figure*}

\begin{figure*}[ht!]
\begin{center}
\includegraphics[scale=0.45]{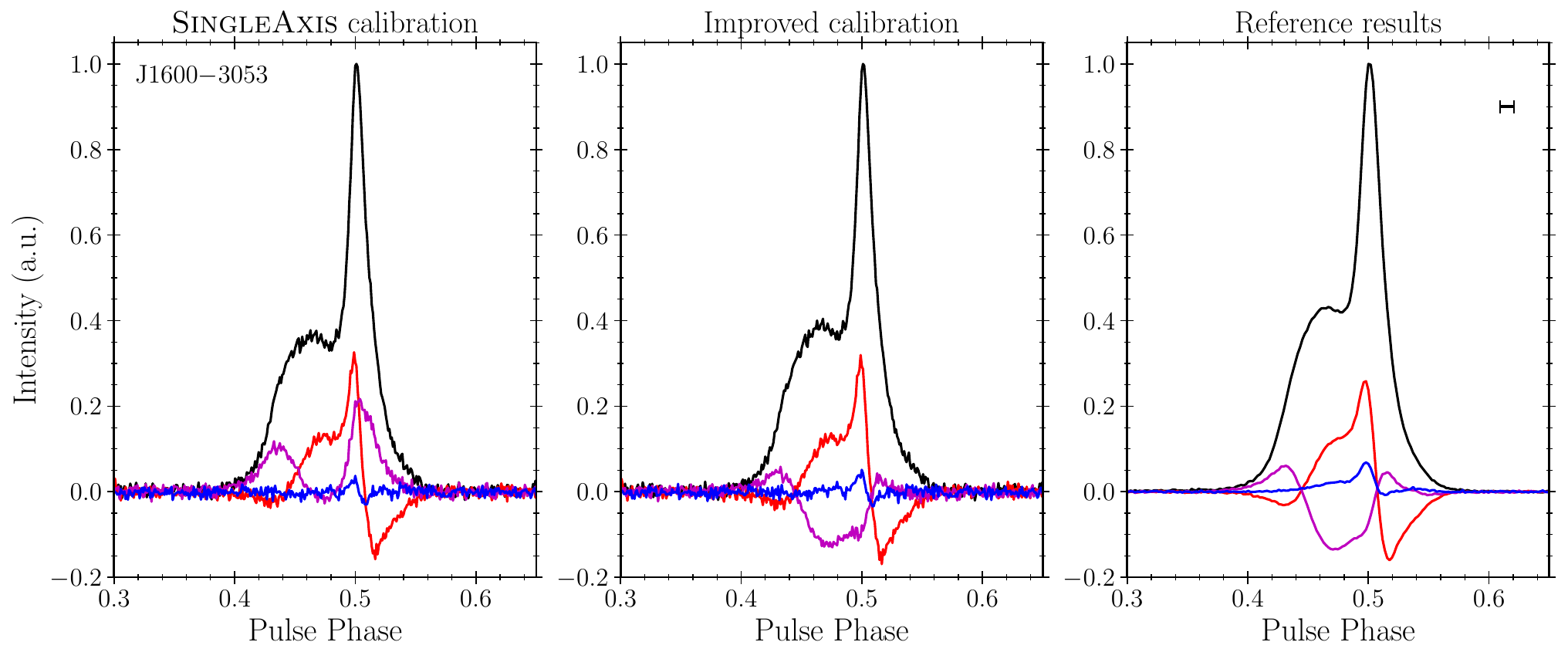}
\includegraphics[scale=0.45]{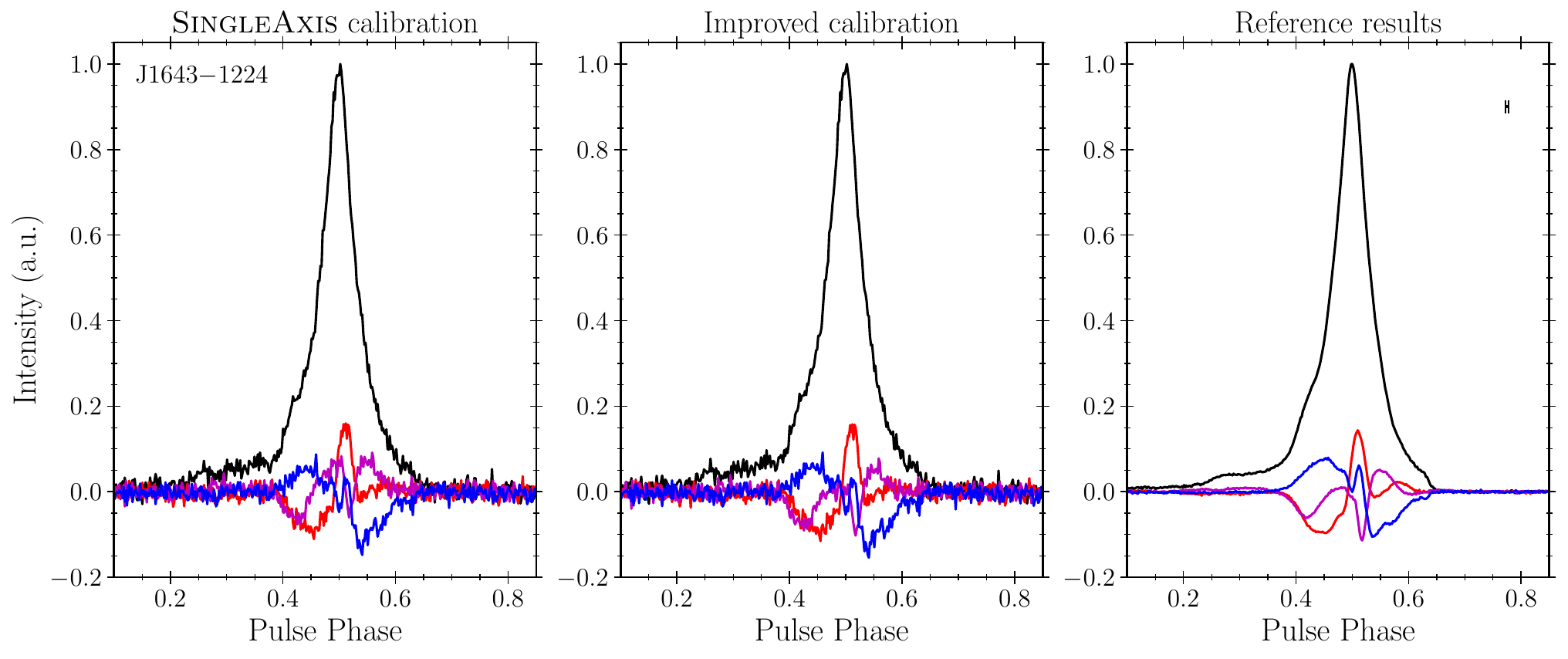}
\includegraphics[scale=0.45]{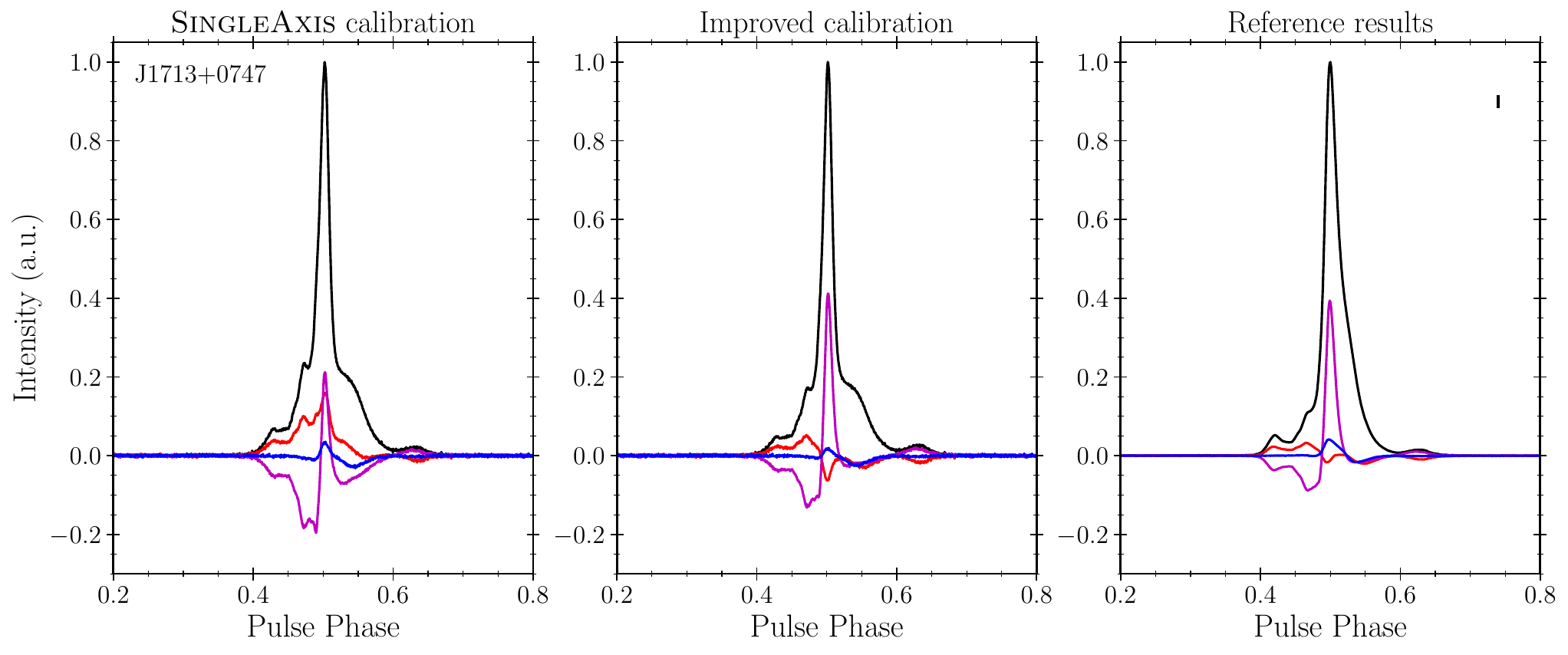}
\caption{Same as Figure~\ref{fig:pol_profs_1}, for PSRs J1600$-$3053, J1643$-$1224, and J1713+0747.}
\label{fig:pol_profs_2}
\end{center}
\end{figure*}

\begin{figure*}[ht!]
\begin{center}
\includegraphics[scale=0.45]{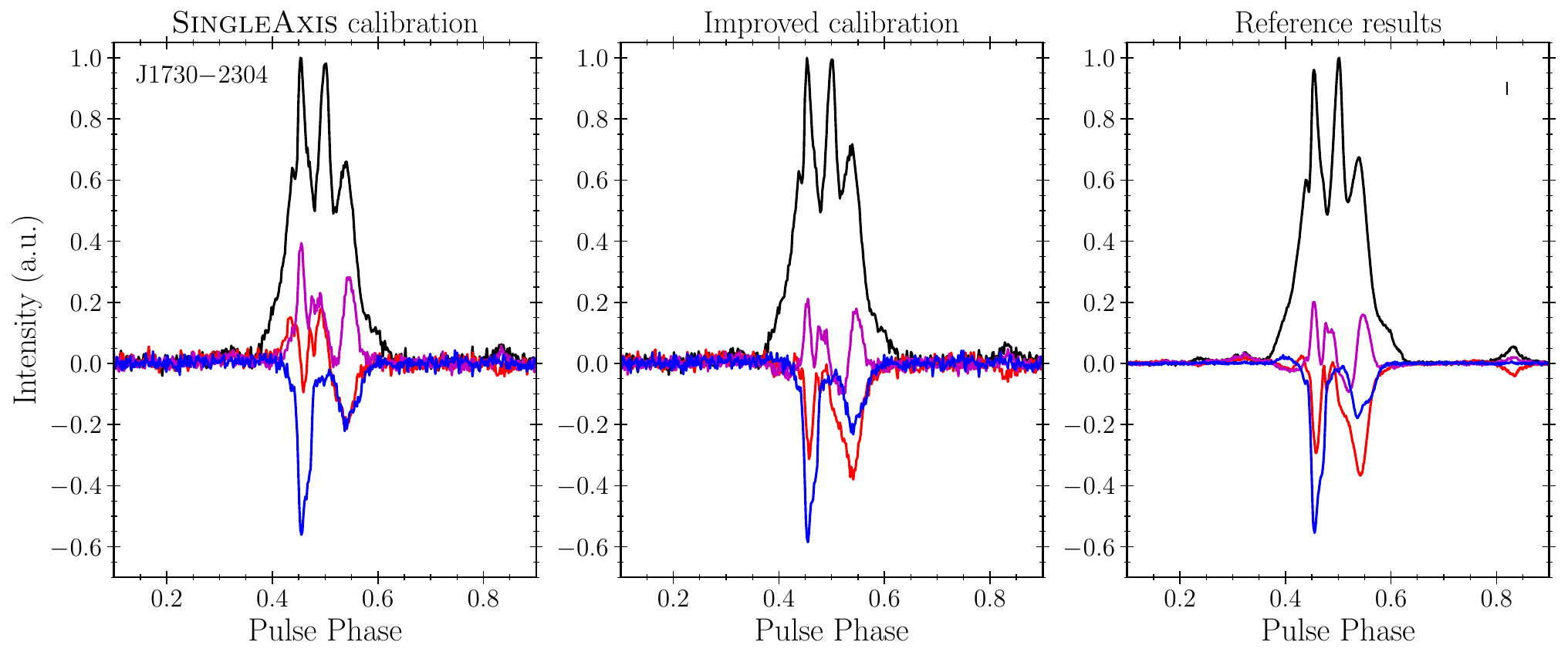}
\includegraphics[scale=0.45]{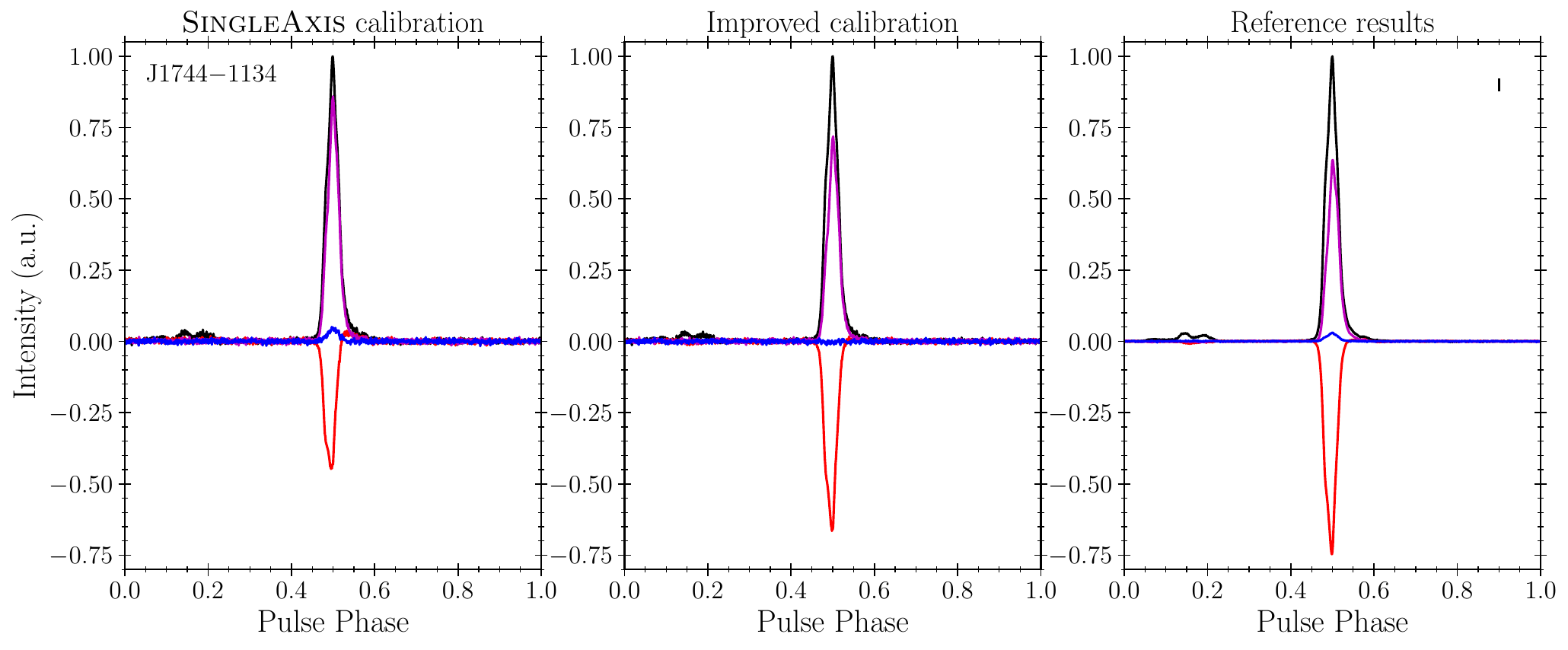}
\includegraphics[scale=0.45]{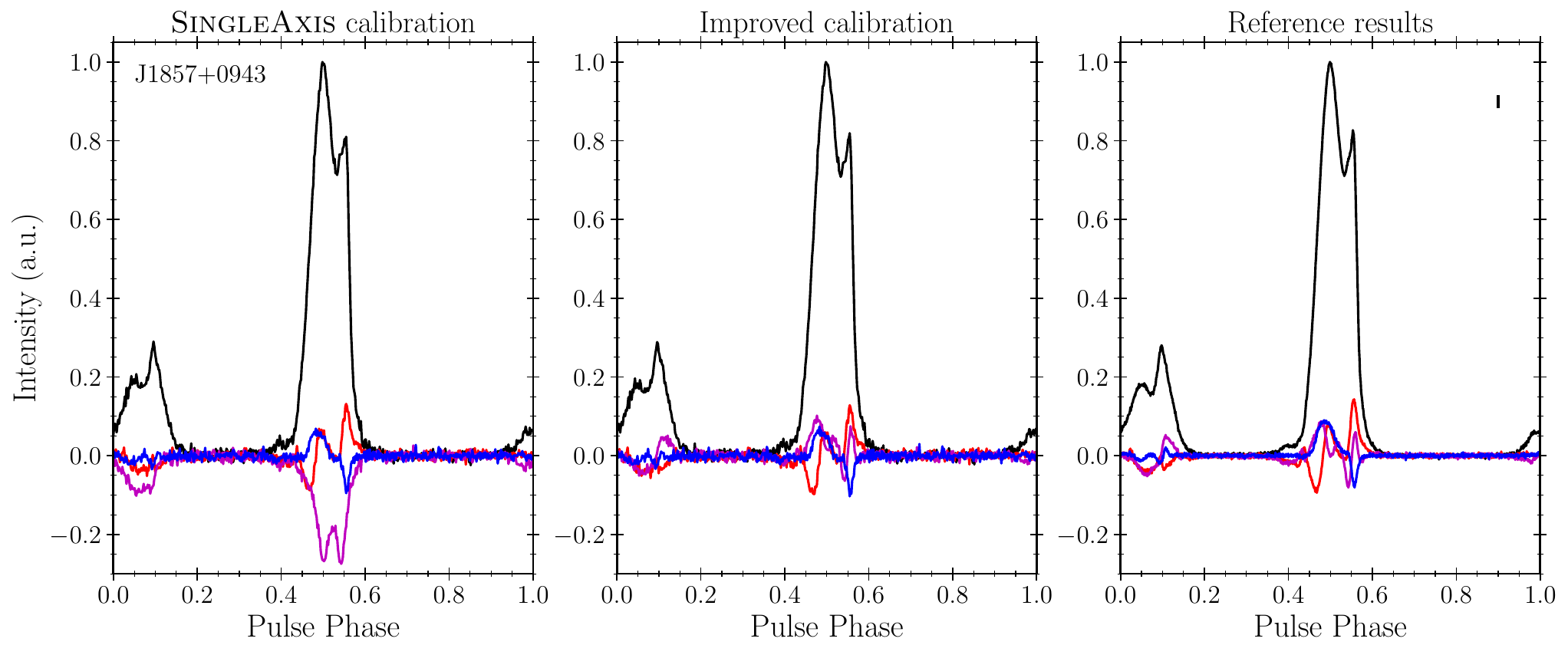}
\caption{Same as Figure~\ref{fig:pol_profs_1}, for PSRs J1730$-$2304, J1744$-$1134, and J1857+0943.}
\label{fig:pol_profs_3}
\end{center}
\end{figure*}

\begin{figure*}[ht!]
\begin{center}
\includegraphics[scale=0.45]{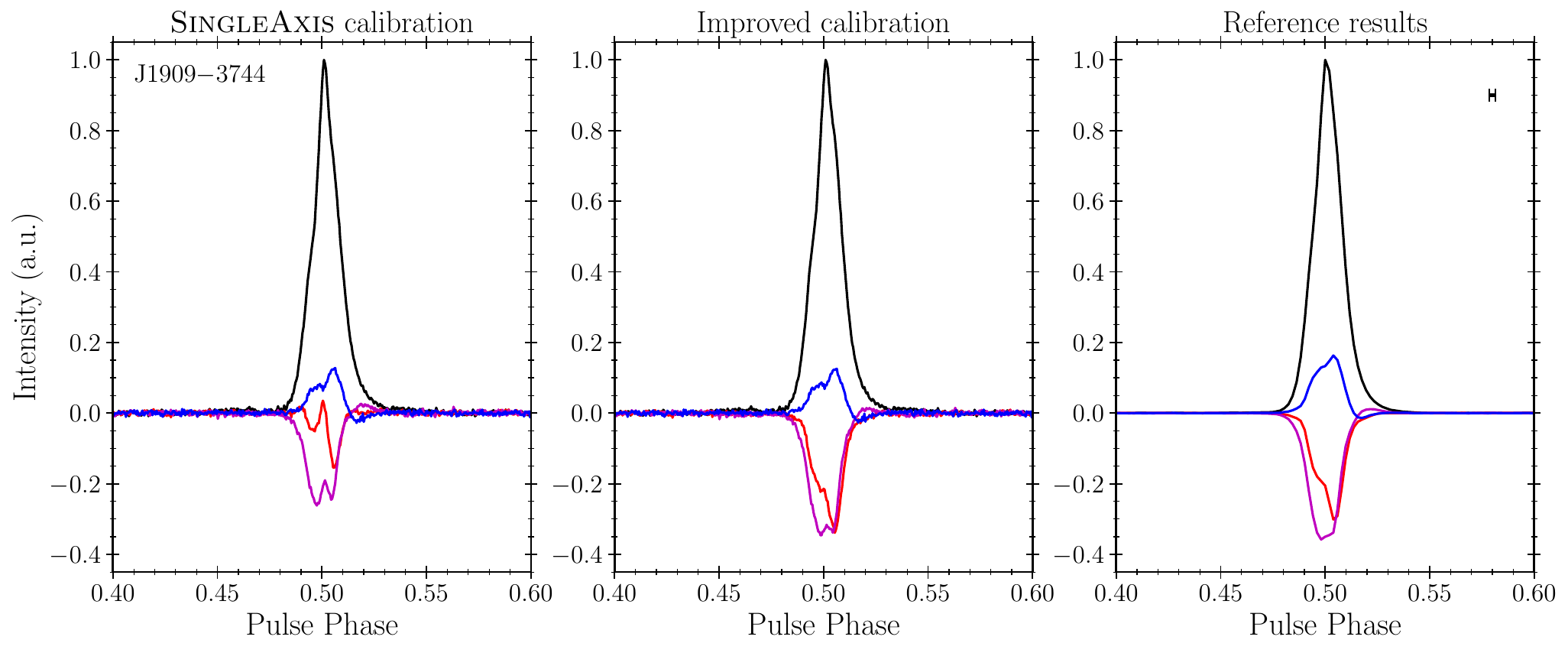}
\includegraphics[scale=0.45]{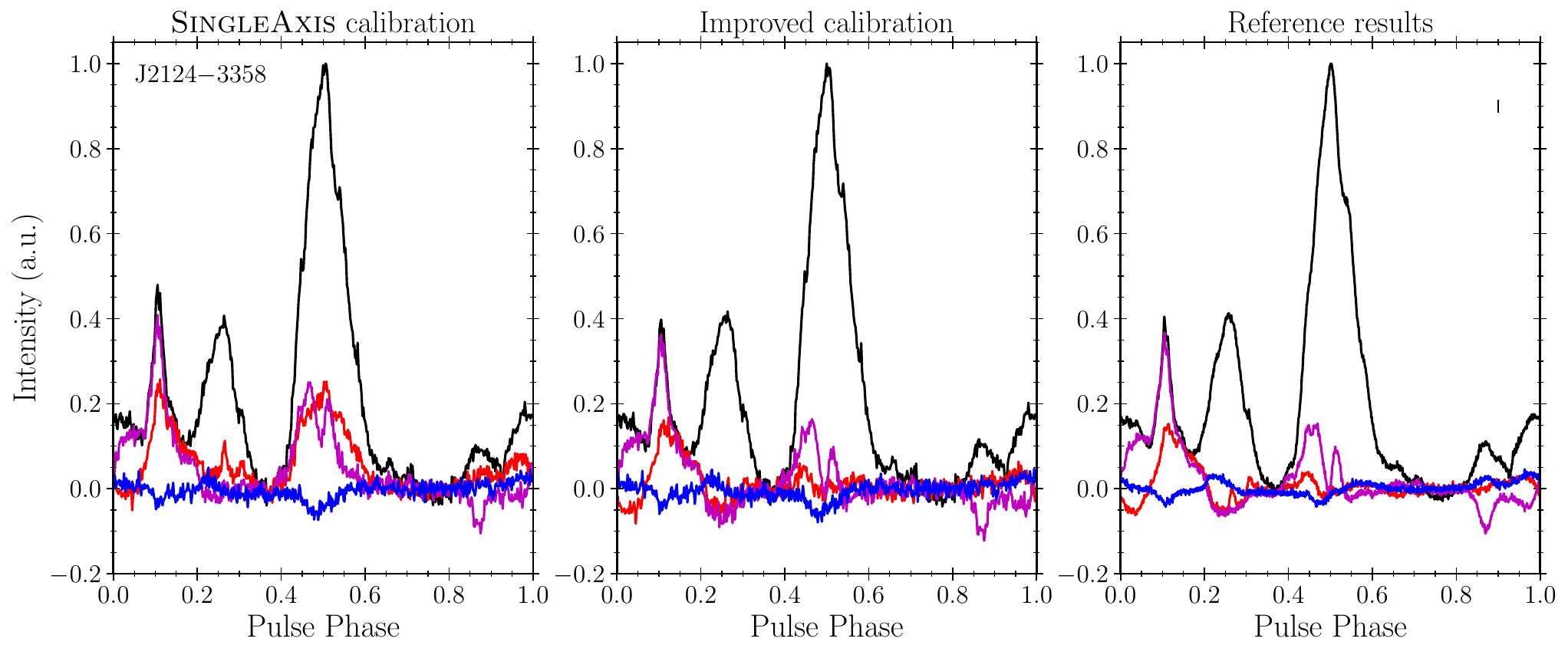}
\includegraphics[scale=0.45]{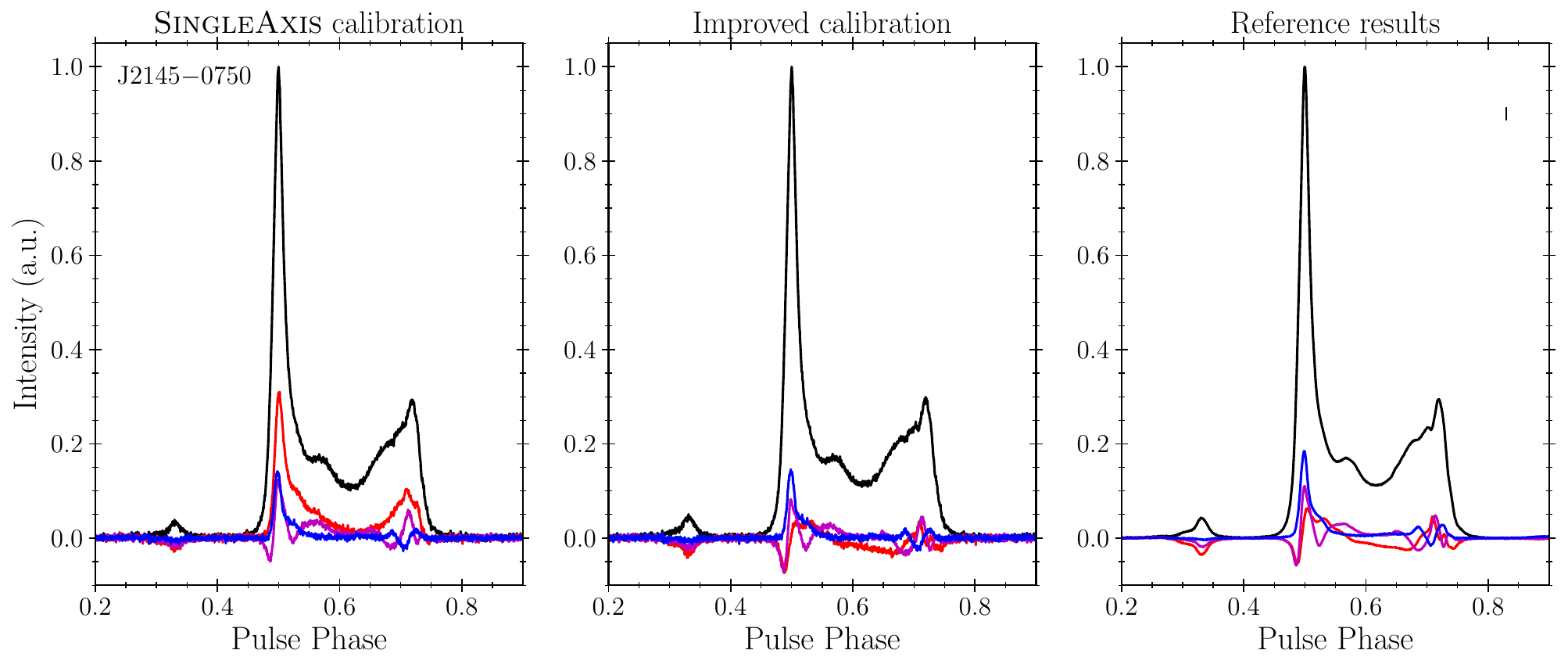}
\caption{Same as Figure~\ref{fig:pol_profs_1}, for PSRs J1909$-$3744, J2124$-$3358, and J2145$-$0750.}
\label{fig:pol_profs_4}
\end{center}
\end{figure*}

\begin{figure*}
\begin{center}
\includegraphics[width=0.9\columnwidth]{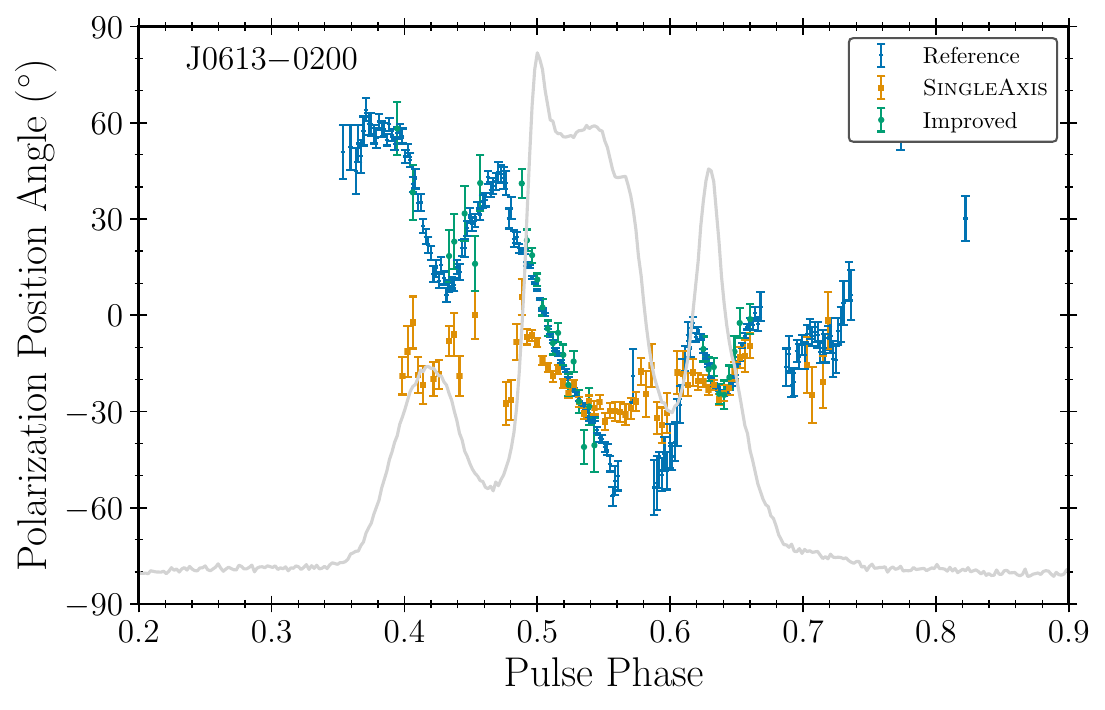}
\includegraphics[width=0.9\columnwidth]{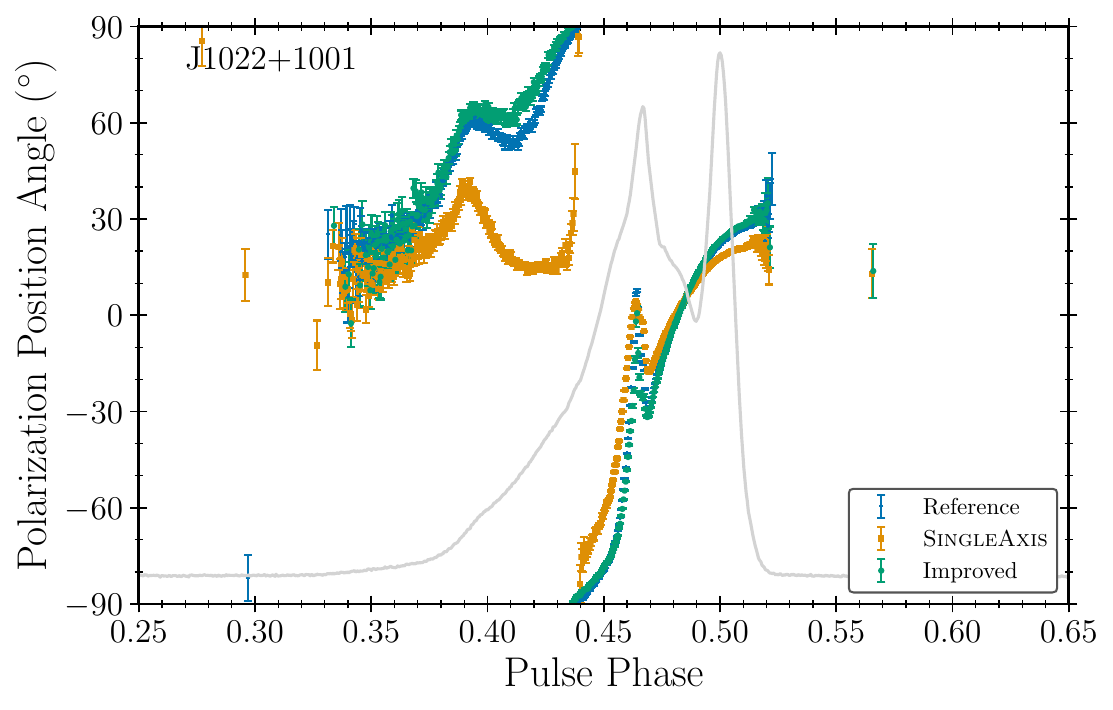}

\includegraphics[width=0.9\columnwidth]{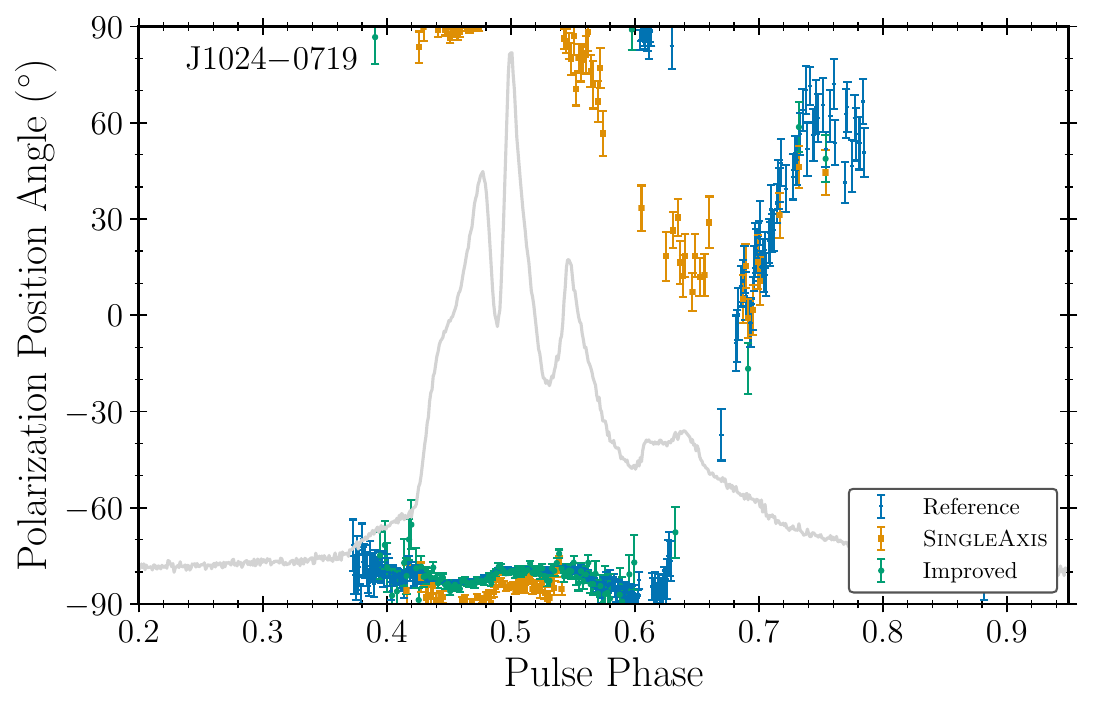}
\includegraphics[width=0.9\columnwidth]{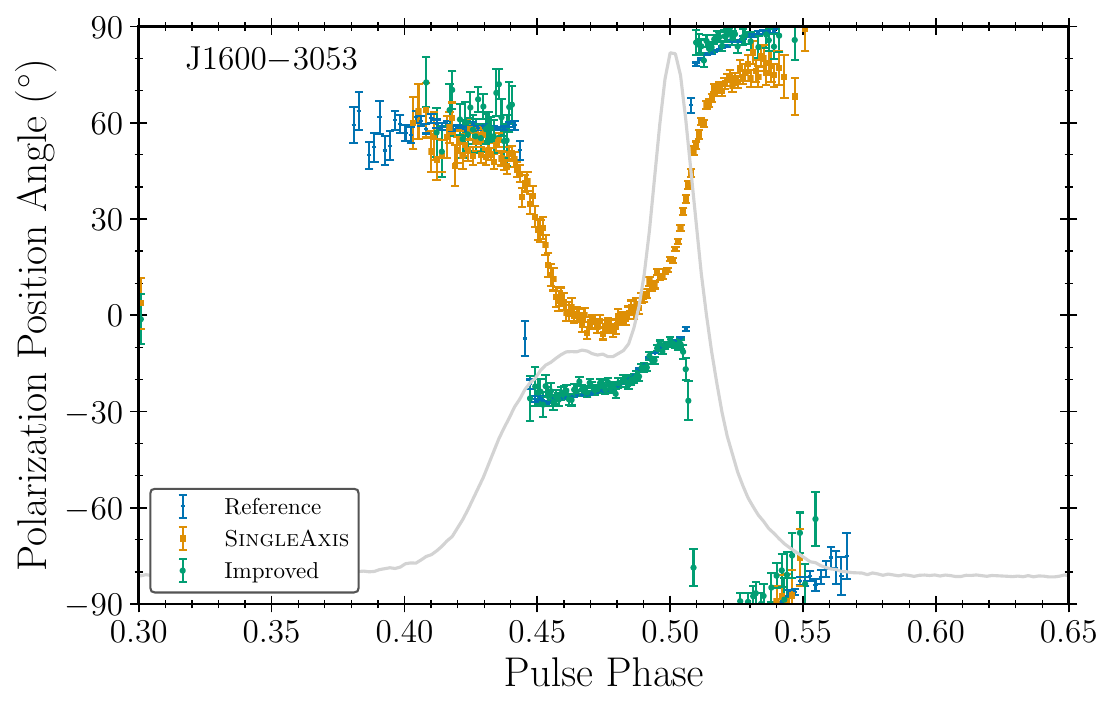}

\includegraphics[width=0.9\columnwidth]{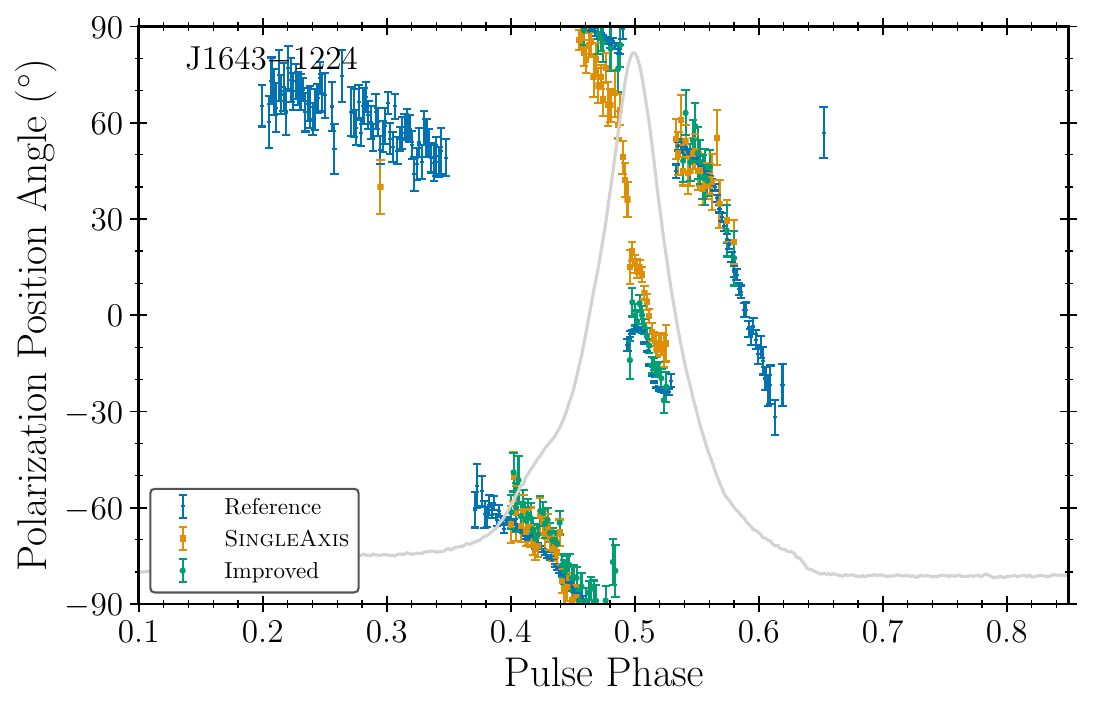}
\includegraphics[width=0.9\columnwidth]{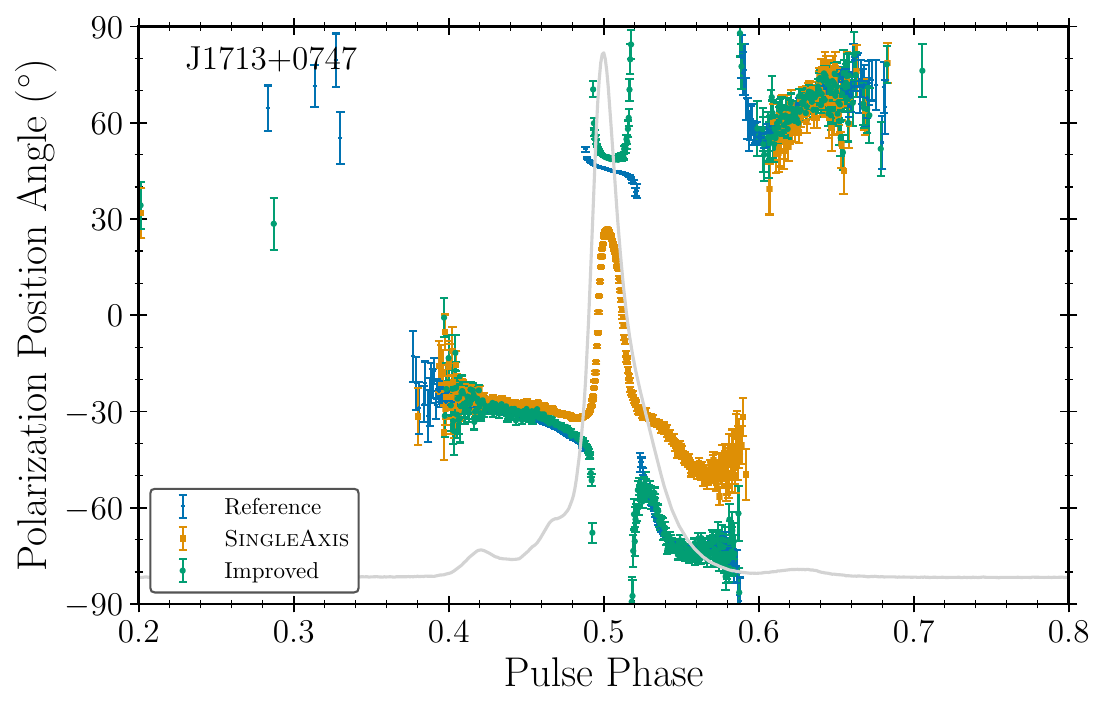}
\caption{Position angle of the linear polarization as a function of pulse phase, for PSRs J0613$-$0200, J1022+1001, J1024$-$0719, J1600$-$3053, J1643$-$1224, and J1713+0747. The position angles, $\Psi$, were calculated using the $Q$ and $U$ Stokes parameters plotted in Figures \ref{fig:pol_profs_1} to \ref{fig:pol_profs_4}, as $\Psi = \frac{1}{2} \arctan{\left(\frac{U}{Q}\right)}$. Results from NUPPI observations calibrated with the \textsc{SingleAxis} method are represented as orange squares. Those obtained with the improved calibration scheme are shown as green circles. Reference results from \citet{Dai2015} are shown as blue crosses. For each pulsar, the pulse profile is shown in light gray for reference.}
\label{fig:PAs_1}
\end{center}
\end{figure*}

\begin{figure*}
\begin{center}
\includegraphics[width=0.95\columnwidth]{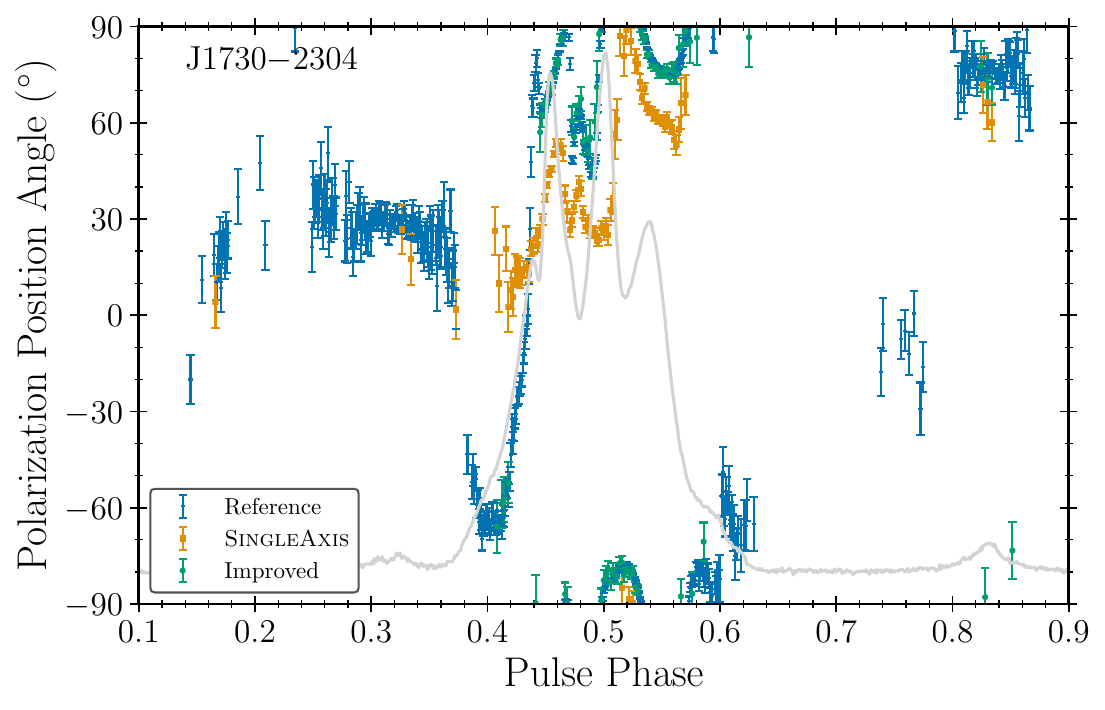}
\includegraphics[width=0.95\columnwidth]{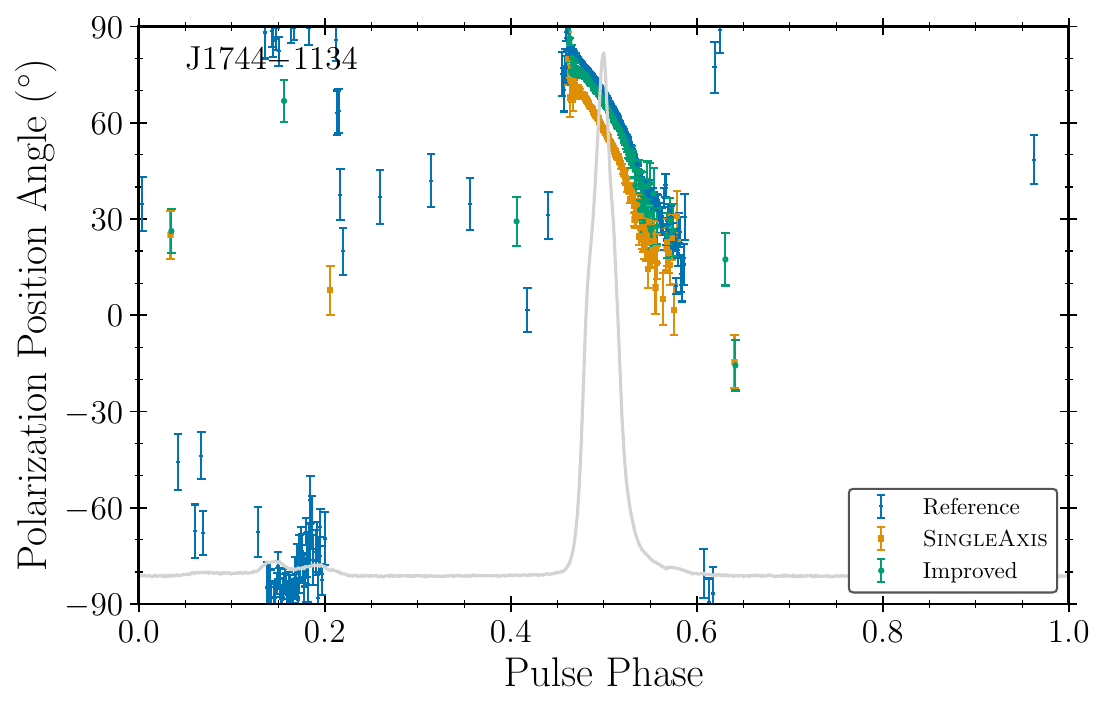}

\includegraphics[width=0.95\columnwidth]{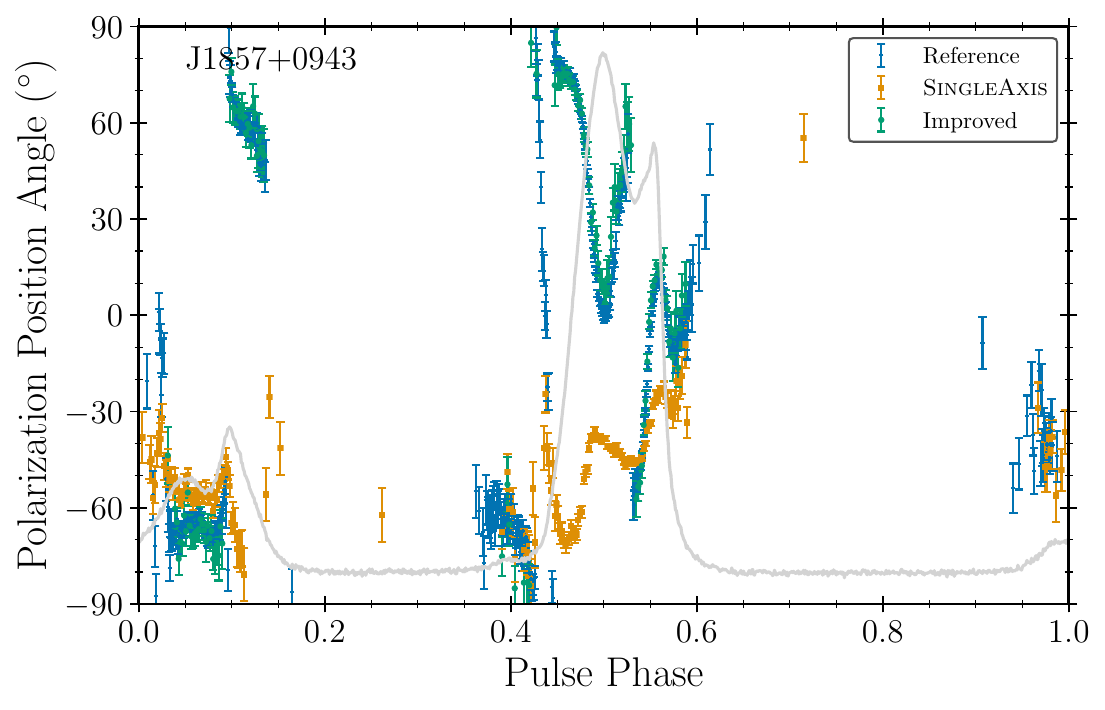}
\includegraphics[width=0.95\columnwidth]{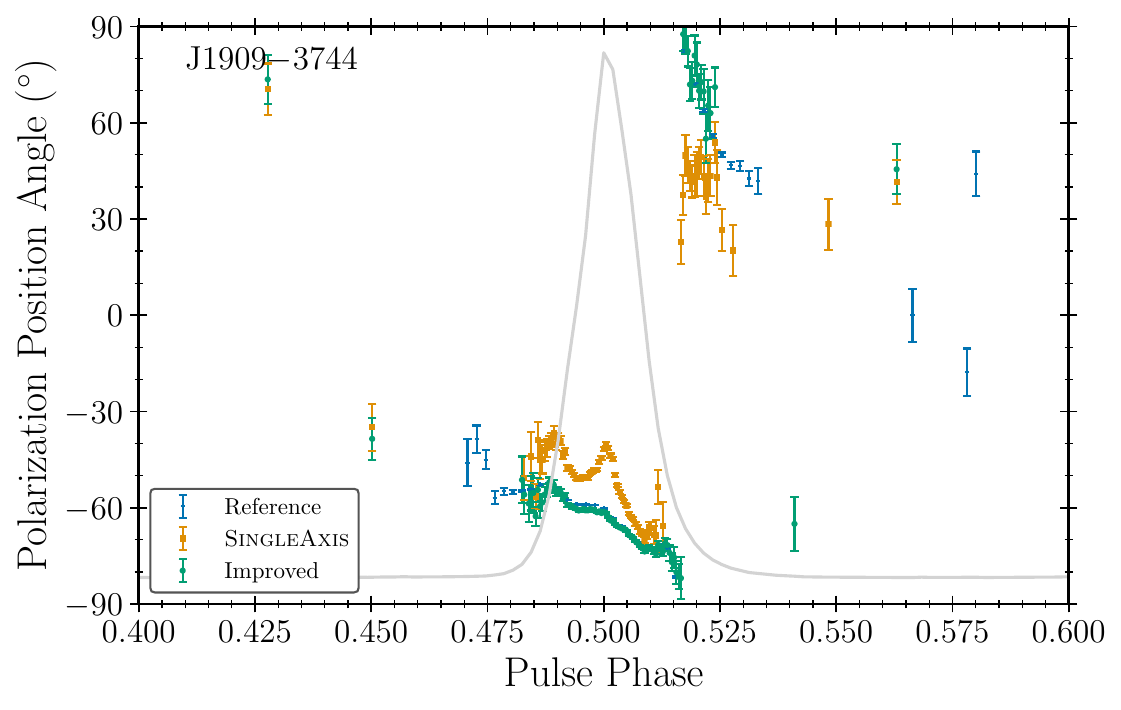}

\includegraphics[width=0.95\columnwidth]{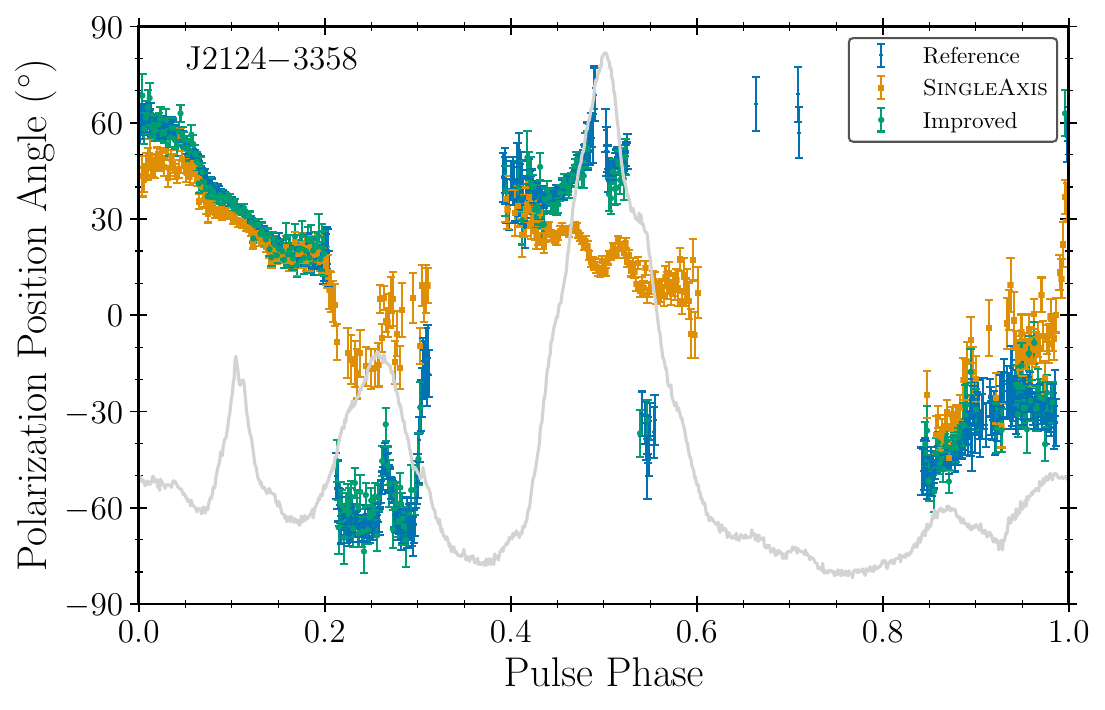}
\includegraphics[width=0.95\columnwidth]{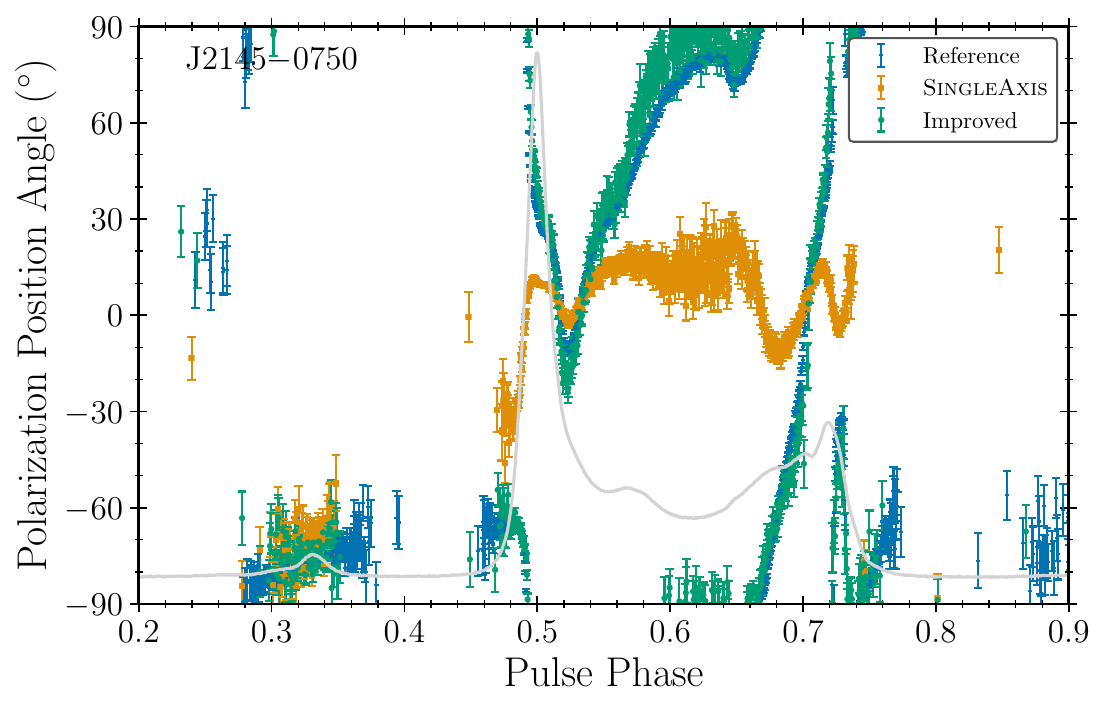}
\caption{Same as Figure~\ref{fig:PAs_1}, for PSRs J1730$-$2304, J1744$-$1134, J1857+0943, J1909$-$3744, J2124$-$3358, and J2145$-$0750.}
\label{fig:PAs_2}
\end{center}
\end{figure*}

In the left-hand panels of Figures~\ref{fig:pol_profs_1} to \ref{fig:pol_profs_4} we show polarimetric profiles at 1.4 GHz for a selection of MSPs observed with NUPPI and calibrated using the \textsc{SingleAxis} method of PSRCHIVE. For each pulsar we selected a high S/N observation, which was then calibrated using the first order calibration parameters (gains, differential gains, and differential phases) determined from the analysis of the noise diode observation conducted prior to it. The data were cleaned of radio-frequency interference (RFI) using the \textsc{Surgical} method of the \textsc{CoastGuard} pulsar analysis library \citep{Lazarus2016}. The original number of 2048 profile bins was reduced by a factor of up to eight to improve readability, and the data were corrected for Faraday rotation across the band caused by magnetic fields along the line-of-sight to the pulsars, using the Rotation Measure (RM) values from \citet{Dai2015}. The average baseline levels for the $I$, $Q$, $U$, and $V$ Stokes parameters were set to zero. Stokes parameters in the figures are in accordance with the conventions described in \citet{vanStraten2010}. The average parallactic angles of all the considered observations were within $\pm 1^\circ$, except for that of J1730$-$2304, for which the average parallactic angle was $-1.8^\circ$. Additional details of the selected observations can be found in Table~\ref{tab:obstable}. The right-hand panels of Figures~\ref{fig:pol_profs_1} to \ref{fig:pol_profs_4} show ``reference'' polarimetric profiles from \citet{Dai2015}, obtained by analyzing 20~cm data from the Parkes telescope in Australia. These profiles were obtained by summing up multiple observations conducted at a central frequency of 1.369~GHz, and over a bandwidth of 256~MHz, \textit{i.e.,} from 1.241 to 1.497~GHz. We discarded NUPPI frequency channels outside of this frequency range to prevent potential pulse profile evolution with frequency from altering the integrated NUPPI profiles, and therefore facilitate a comparison with the \citet{Dai2015} results. Unlike NUPPI data, the 20~cm Parkes data analyzed in \citet{Dai2015} were not coherently dedispersed. As a result, the profiles are affected by dispersion smearing, indicated by markers at the top right. Details on dispersion smearing and the amount of smearing for each of the considered pulsars are given in \citet{Dai2015}. In a few cases, such as those of PSRs~J0613$-$0200, J1600$-$3053, and J1909$-$3744, narrow profile structures visible in the NUPPI profiles can be smeared out in the Parkes profiles, due to this effect. Finally, in Figures~\ref{fig:PAs_1} and \ref{fig:PAs_2} we show plots of the position angles (PAs), $\Psi$, of the linear polarization as a function of rotational phase for the same pulsars, calculated as: 

\begin{equation}
\Psi = \dfrac{1}{2} \arctan \left( \dfrac{U}{Q} \right).
\end{equation}

We note that the PAs shown in Figures~\ref{fig:PAs_1} and \ref{fig:PAs_2} were obtained by correcting Faraday rotation to infinite frequency\footnote{With the ``\texttt{-{}-aux\_rm}'' option of the PSRCHIVE tool ``\texttt{pam}''.}, hence the fact that the ``reference'' PAs are different from those shown in \citet{Dai2015}.

\begin{table*}
\begin{small}
\caption[]{Properties of the 1.4 GHz NUPPI observations presented in Figures~\ref{fig:pol_profs_1} to \ref{fig:pol_profs_4}. The first five columns give the pulsar names and discovery articles, their rotational periods (P), and Dispersion Measure (DM) values, and the Rotation Measure (RM) values from \citet{Dai2015} used for correcting the NUPPI data for Faraday rotation. The following columns give the observation epochs and durations, the numbers of pulse phase bins of the NUPPI profiles displayed in Figures~\ref{fig:pol_profs_1} to \ref{fig:pol_profs_4}, and the signal-to-noise (S/N) ratios of the profiles, when calibrating the data using the \textsc{SingleAxis} method of PSRCHIVE (``Orig. S/N'') and with the improved calibration method presented in Section~\ref{subsec:calib2} (``New S/N''). See Section~\ref{subsec:calib2} and Equation~\ref{eq:snr} for details on the calculation of the S/N values.}
\label{tab:obstable}
\centering

\begin{tabular}{ccccccccccc}
\hline
\hline
Pulsar & Discovery article & P & DM & RM & Epoch & Duration & $N_\mathrm{bins}$ & Orig. S/N & New S/N \\
 & & (ms) & (pc cm$^{-3}$) & (rad m$^{-2}$) & (MJD) & (s) & & & \\
\hline
J0613$-$0200 & \citet{disc0613_1643_1730} & 3.06 & 38.77 & 9.7 & 59346 & 3192.3 & 256 & 279.1 & 316.6 (+13.4\%) \\
J1022+1001 & \citet{disc1022} & 16.45 & 10.27 & $-$0.6 & 59339 & 3931.7 & 2048 & 6898.1 & 6384.0 ($-$7.5\%) \\
J1024$-$0719 & \citet{disc1024_1744_2124} & 5.16 & 6.49 & $-$8.2 & 59323 & 3901.1 & 512 & 349.1 & 454.6 (+30.2\%) \\
J1600$-$3053 & \citet{disc1600} & 3.60 & 52.33 & $-$15.5 & 59338 & 3467.7 & 1024 & 794.8 & 841.5 (+5.9\%) \\
J1643$-$1224 & \citet{disc0613_1643_1730} & 4.62 & 62.40 & $-$308.1 & 59400 & 1897.0 & 512 & 561.1 & 579.3 (+3.2\%) \\
J1713+0747 & \citet{disc1713} & 4.57 & 15.99 & 8.4 & 59384 & 2830.2 & 2048 & 4736.7 & 4966.0 (+4.8\%) \\
J1730$-$2304 & \citet{disc0613_1643_1730} & 8.12 & 9.63 & $-$7.2 & 59368 & 2080.6 & 512 & 484.4 & 509.3 (+5.1\%) \\
J1744$-$1134 & \citet{disc1024_1744_2124} & 4.07 & 3.14 & $-$1.6 & 59347 & 1902.1 & 1024 & 1236.4 & 1510.6 (+22.2\%) \\
J1857+0943 & \citet{disc1857} & 5.36 & 13.30 & 16.4 & 59401 & 3039.3 & 512 & 1025.0 & 1018.3 ($-$0.7\%) \\
J1909$-$3744 & \citet{disc1909} & 2.95 & 10.39 & $-$6.6 & 59359 & 3707.4 & 2048 & 1242.1 & 1418.1 (+14.2\%) \\
J2124$-$3358 & \citet{disc1024_1744_2124} & 4.93 & 4.60 & $-$5.0 & 59365 & 2738.4 & 512 & 439.7 & 425.8 ($-$3.2\%) \\
J2145$-$0750 & \citet{disc2145} & 16.05 & 9.00 & $-$1.3 & 59338 & 3013.8 & 2048 & 3004.3 & 3133.6 (+4.3\%) \\
\hline
\end{tabular}
\end{small}
\end{table*}

It is clear, from Figures~\ref{fig:pol_profs_1} to \ref{fig:PAs_2}, that the polarimetric profiles and linear polarization angles derived from NUPPI data calibrated using the \textsc{SingleAxis} method are highly inconsistent with the reference results from \citet{Dai2015}. The Parkes Multibeam receiver, used to collect the data presented in the latter article, is also equipped with a noise diode injecting a 100\% linearly polarized signal, which is used to determine differential gains and phases prior to the pulsar observations. However, in addition to the noise diode observations, observations of the nearby, highly polarized MSP J0437$-$4715 \citep{Johnston1993} are regularly conducted over wide ranges of parallactic angles in order to determine cross-couplings and polarization leakages between the feeds, as described in \citet{vanStraten2004}. This model, developed further in Section~\ref{subsec:calib2}, provides a more accurate estimate of the instrumental response. Polarimetric profiles and PA measurements from NUPPI data calibrated using the \textsc{SingleAxis} method are therefore incorrect. This polarimetric calibration procedure was used until an improved calibration scheme (see next section) was developed in late 2019.

%%%%%%%%

\subsection{Improving the polarimetric calibration of NUPPI observations}
\label{subsec:calib2}

As demonstrated in Section~\ref{subsec:calib1}, a simple polarimetric calibration of NUPPI data that assumes perfectly orthogonally polarized feeds and an ideal, 100\% linearly polarized noise diode signal illuminating the two feeds equally and in phase does not produce satisfactory results. Cross-coupling between the two feeds, and a potentially non-ideal noise diode signal, can alter polarimetric results. In order to determine these second-order calibration corrections, a number of studies have used observations of bright, broadband-emitting polarized pulsars over wide ranges of parallactic angles \citep[see e.g.,][]{Stinebring1984,Xilouris1991,Johnston2002}. Doing so, different orientations of the angle of polarization of the pulsar are sampled during the observation, and cross-coupling of the polarization feeds can in turn be determined. The NRT, however, can only observe sources for approximatively one hour around transit. From Equation~\ref{eq:parallactic}, it can be seen that in the most favorable situation, corresponding to $\delta = 90^\circ$, the parallactic angle varies by only $15^\circ$ in an hour of observation, insufficient for appropriately sampling pulsar polarization angles. However, the discrepancies between NUPPI polarimetric profiles obtained with a simple \textsc{SingleAxis} calibration with published results (such as the comparisons made in Section~\ref{subsec:calib1}) prompted us to explore strategies for circumventing the above-mentioned limitation of the telescope. The solution found in late 2019 is to conduct observations of bright, polarized pulsars in a special observational mode, in which the feed horn, which is held fixed during normal observations, is made to rotate by 180$^\circ$ during the $\sim 1$-hr observation, \textit{i.e.,} at a rate of $\sim 3^\circ$ per minute, close to the maximum horn rotation rate allowed by the instrument. Doing so effectively mimics a variation of the parallactic angle, thereby enabling the NRT to sample much wider ranges of pulsar polarization angles during the observation.

The framework for simultaneously determining the full polarimetric response of the receiver and the actual Stokes parameters of the reference signal from the analysis of such observations over wide parallactic angle ranges, or, as in our case, over wide horn orientation ranges, is presented in \citet{vanStraten2004} and is implemented in the \textsc{Reception} calibration model of PSRCHIVE. In this model, the Jones matrix representing the instrumental response is given by

\begin{equation}
\boldsymbol{J}_\mathrm{R} = G\ \boldsymbol{B}_{\boldsymbol{\hat{q}}} \left(\gamma\right)\ \boldsymbol{R}_{\boldsymbol{\hat{q}}} \left(\varphi\right)\ \boldsymbol{C}.
\label{eq:jones_reception}
\end{equation}

In this equation, the term $\boldsymbol{C}$, which represents the response of a receiver with non-ideal feeds, is given by

\begin{equation}
\boldsymbol{C} = \boldsymbol{\delta}_0 \ \boldsymbol{S} \left(\theta_0,\ \epsilon_0\right) + \boldsymbol{\delta}_1 \ \boldsymbol{S} \left(\theta_1,\ \epsilon_1\right),
\label{eq:c}
\end{equation}

\noindent
where $\boldsymbol{\delta}_i$ is the 2 $\times$ 2 selection matrix: 

\begin{equation}
\boldsymbol{\delta}_i = \begin{pmatrix}
\delta_{0i} & 0 \\
0 & \delta_{1i} \\ 
\end{pmatrix},
\end{equation}

\noindent
with $\delta_{ij}$ the Kronecker delta. In Equation~\ref{eq:c}, $\epsilon$ and $\theta$ represent the ellipticities and orientations of the two feeds, and $\boldsymbol{S} \left(\theta,\ \epsilon\right) = \boldsymbol{R}_{\boldsymbol{\hat{u}}} \left(\epsilon \right)\ \boldsymbol{R}_{\boldsymbol{\hat{v}}} \left(\theta \right)$. The model describing the noise diode observation is then $\boldsymbol{\rho^\prime}_\mathrm{ref} = \boldsymbol{J}_\mathrm{R}\ \boldsymbol{\rho}_\mathrm{ref} \ \boldsymbol{J}_\mathrm{R}^\dagger$, where $\boldsymbol{\rho}_\mathrm{ref}$ no longer corresponds to the ideal reference signal given in Equation~\ref{eq:ideal_diode}, and the model representing the pulsar observation is 

\begin{equation}
\boldsymbol{\rho}_\mathrm{psr}^\prime (\Phi^\prime) = \boldsymbol{J}_\mathrm{R}\ \boldsymbol{R}_{\boldsymbol{\hat{v}}} (\Phi^\prime)\ \boldsymbol{\rho}_\mathrm{psr}\ \boldsymbol{R}_{\boldsymbol{\hat{v}}}^{\boldsymbol{\dagger}} (\Phi^\prime) \boldsymbol{J}_\mathrm{R}^\dagger,
\label{eq:rotation}
\end{equation}

\noindent
with $\Phi^\prime = \Phi + \alpha$ where $\alpha$ is the orientation of the horn at a given time, and $\boldsymbol{\rho}_\mathrm{psr}$ is the coherency matrix that is intrinsic to the observed pulsar signal. Given $\boldsymbol{\rho^\prime}_\mathrm{ref}$ and $\boldsymbol{\rho^\prime}_\mathrm{psr}$ for different values of $\Phi^\prime$, and assuming $\boldsymbol{\rho^\prime}_\mathrm{psr}$ is constant in all time samples, the \texttt{Reception} model can solve for the actual Stokes parameters of the reference noise diode signal, and for the values of $G$, $\gamma$, and $\varphi$ and of the ellipticies, $\epsilon$, and orientations, $\theta$, of the two feeds. In practice, multiple on-pulse phase bins from a given pulsar can be included as input source polarizations, to increase the number of constraints for the fit. Finally, as described in \citet{Ord2004}, the polarimetric response determined from the above-described analysis at a reference epoch can be used to calibrate regular pulsar observations using the following Jones matrix: 

\begin{equation}
\boldsymbol{J} = G^\prime \ \boldsymbol{B}_{\boldsymbol{\hat{q}}} \left(\gamma^\prime\right)\ \boldsymbol{R}_{\boldsymbol{\hat{q}}} \left(\varphi^\prime\right)\ \boldsymbol{J}_0 . 
\label{eq:calibration}
\end{equation}

In the above expression, $\boldsymbol{J}_0$ is the Jones matrix representing the instrumental response at the reference epoch, and $G^\prime$, $\gamma^\prime$, and $\varphi^\prime$ denote the changes in absolute gain, differential gain, and differential phase since that reference epoch, as determined from a \texttt{SingleAxis} analysis of a new observation of the noise diode, using the actual Stokes parameters of the diode determined from the \texttt{Reception} analysis. This assumes that the polarization properties of the noise diode have not changed since the reference epoch, and similarly, the model described in Equation~\ref{eq:calibration} assumes that the ellipticities and orientations also have not varied. 

We conducted the first observation of a bright polarized pulsar with feed horn rotation on MJD~58816 (November 29, 2019). The observed pulsar was PSR~J0742$-$2822 \citep[B0740$-$28,][]{Bonsignori1973}, a relatively bright \citep[flux density at 1.4~GHz of $26 \pm 2$ mJy, see][]{Jankowski2018}, slow-period ($P \sim 0.167$ s) pulsar with a very strong linear polarization component: \citet{Johnston2018} measured a percentage of linear polarization ($L/I$) of $\sim 90$\% for this pulsar, making it an ideal target for our calibration observations. Since MJD~58816, PSR~J0742$-$2822 has been observed with NUPPI in calibration mode on a regular basis, with one to two observations per month on average. We here present the results of the analysis of a calibration mode observation of PSR~J0742$-$2822 conducted on MJD~59368 (June 3, 2021), close in time to the regular mode NUPPI observations presented in Figures~\ref{fig:pol_profs_1} to \ref{fig:pol_profs_4}. We integrated the $\sim 1$-hr observation in time to form one sub-integration per minute of observation. The original frequency resolution of 128 channels of 4~MHz each was kept. The original number of phase bins per rotation of 2048 was reduced by a factor of four in order to increase the S/N of the pulsar in the individual on-pulse phase bins and time sub-integrations. We then used the PSRCHIVE tool ``\texttt{pcm}'' to carry out the \texttt{Reception} model analysis of the observation of PSR~J0742$-$2822 and of the noise diode conducted prior to the pulsar observation, selecting 32 on-pulse phase bins as sources of input pulsar polarization data for the analysis. As explained in detail in Appendix~B of \citet{vanStraten2004}, solutions to Equation~\ref{eq:rotation} are degenerate under commutation, and two assumptions need to be made to constrain the mixing between Stokes $I$ and $V$ on the one hand and between $Q$ and $U$ on the other hand. In our analysis we first assumed that the feeds have equal ellipticities, which implies that the mixing between $I$ and $V$ is zero. In that case, the circular polarization of the reference signal from the noise diode is allowed to vary in the fit. Second, we assumed that the misorientation of the first polarization feed is zero. Under this hypothesis, the $Q$ parameter of the noise diode signal is also allowed to vary in the fit. Finally, the data were normalized by the total invariant interval \citep{Britton2000} in order to limit apparent variations of the gain caused by scintillation of the pulsar across the observation.

\begin{figure*}[ht]
\begin{center}
\includegraphics[scale=0.5]{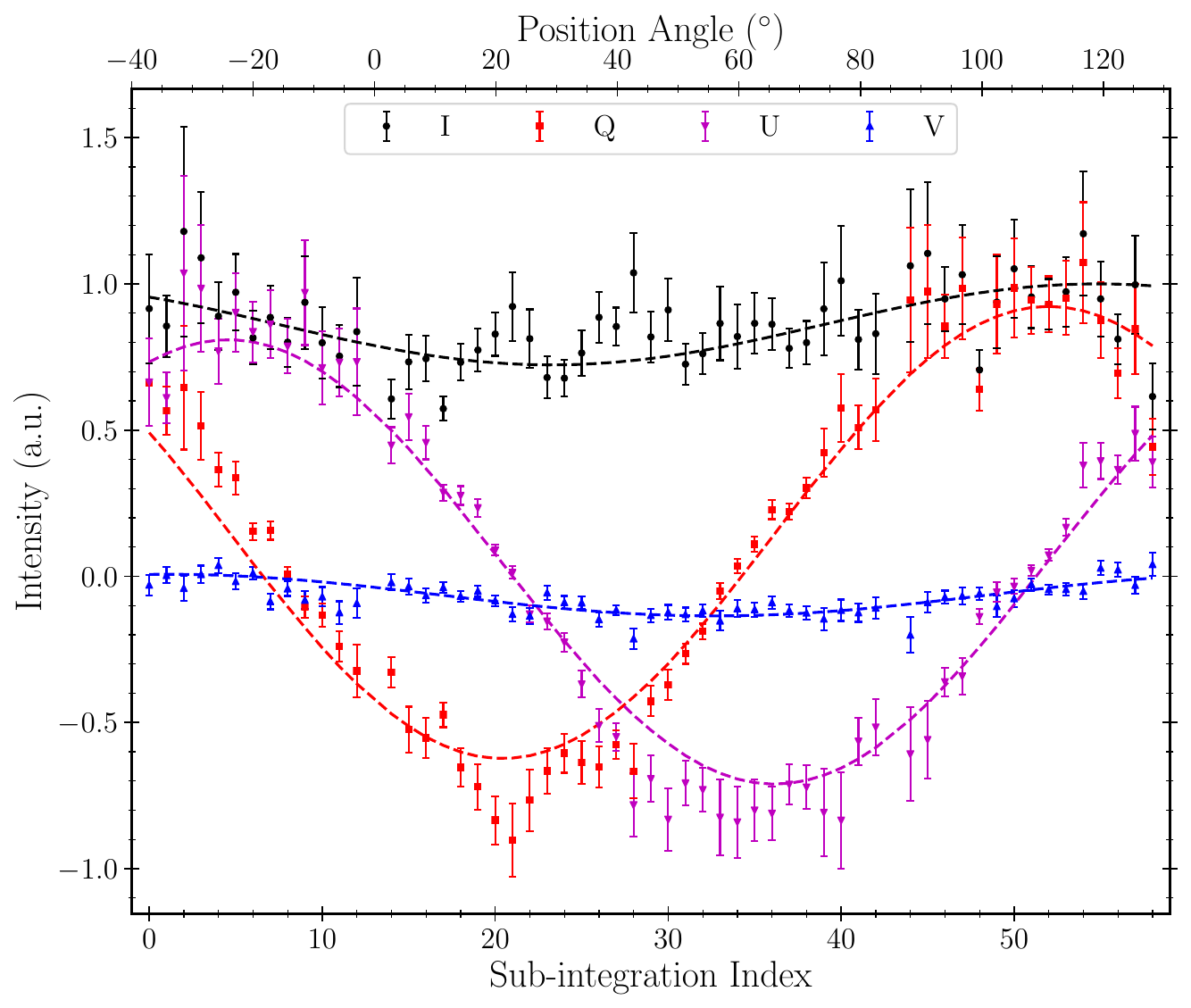}
\caption{Measured and modeled Stokes parameters for PSR~J0742$-$2822 as a function of time and position angle $\Phi^\prime = \Phi + \alpha$, where $\Phi$ is the parallactic angle and $\alpha$ is the orientation of the horn at a given time. The pulsar was observed at 1484~MHz with the NUPPI backend, in an observing mode where the feed horn is made to rotate by $3^\circ$ per minute, during the $\sim$1-hr observation. The modeled Stokes parameters, represented as dashed lines, were determined by analyzing the data with the \texttt{Reception} model of PSRCHIVE, enabling the determination of the polarimetric response of the telescope. The data points shown in this plot correspond to the Stokes parameters in the phase bin in which the pulsar is brightest, with 512 bins per rotation in this case. See Section~\ref{subsec:calib2} for the description of the analysis.}
\label{fig:rot_pcm}
\end{center}
\end{figure*}

In Figure~\ref{fig:rot_pcm} we show a plot of the Stokes parameters as a function of time for the MJD~58816 observation of J0742$-$2822, along with the modeled Stokes parameters represented as dashed lines. The data shown in this figure were taken from the phase bin in which the pulsar is brightest. Stokes parameters $Q$ and $U$, which describe the linear component of the polarized signal, display clear sinusoidal variations, as expected given that the feed horn rotated during the observation. The modeled Stokes parameters generally represent the actual measurements well, with slight discrepancies (particularly in the $Q$ and $U$ parameters) observed during the first half of the observation. These discrepancies, which suggest possible time-dependence of the instrumental response during observations, will be investigated further in future work.

\begin{figure}[ht]
\begin{center}
\includegraphics[width=0.95\columnwidth]{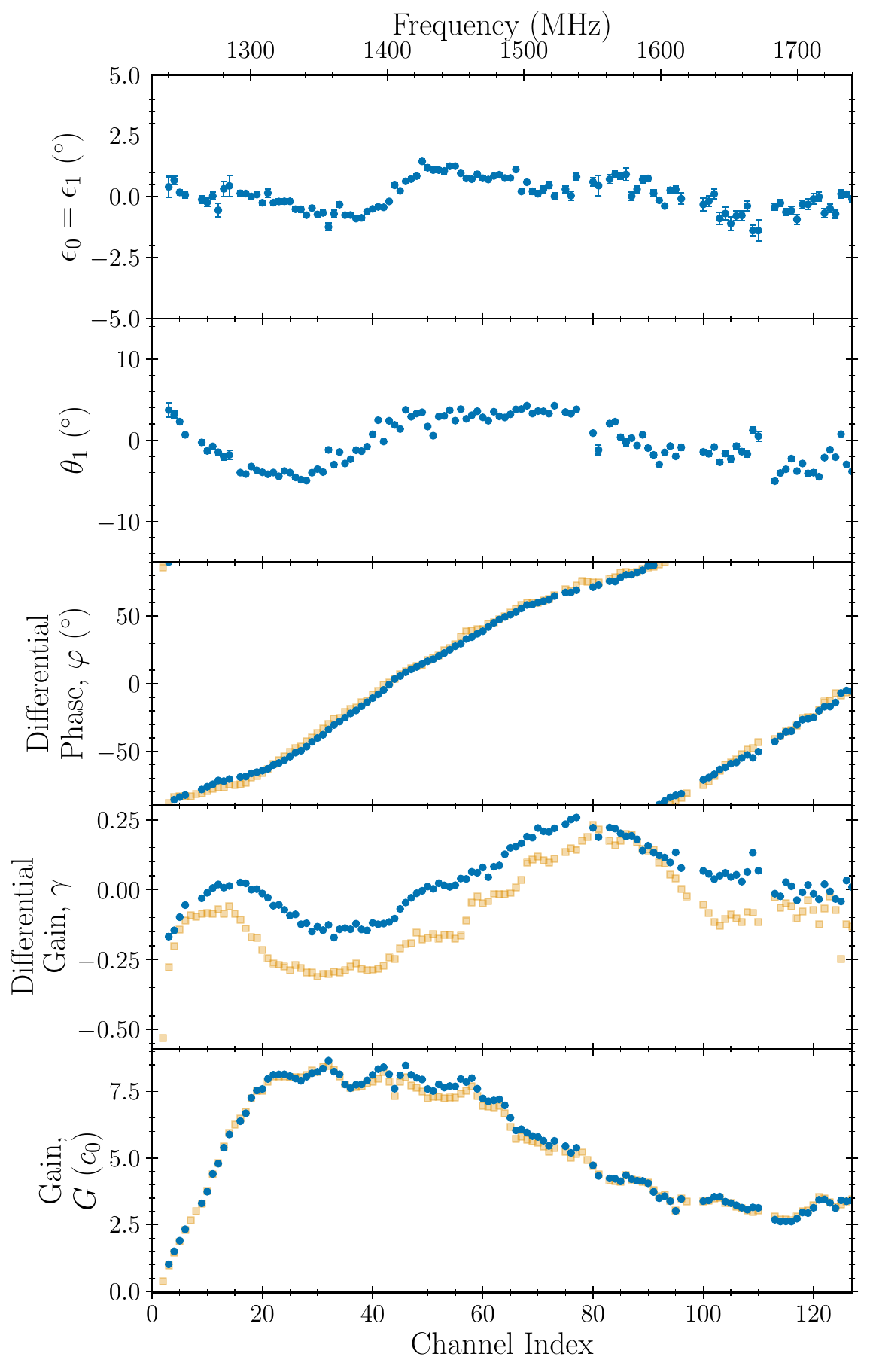}
\caption{Best-fit calibration parameters as determined from a joint analysis of an observation of PSR~J0742$-$2822 with the feed horn rotating by 180$^\circ$ across the $\sim 1$-hr observation, and of an observation of the noise diode conducted prior to the pulsar observation. The three bottom panels are the same as in Figure~\ref{fig:sol_0742}. The blue points represent the best-fit results from this analysis, and the yellow squares show the results from the \texttt{SingleAxis} analysis of the noise diode observation only, as in Figure~\ref{fig:sol_0742}, for comparison. The two top panels show the feed ellipticities (assumed to be equal in our analysis, see Section~\ref{subsec:calib2} for details), and the orientations ($\theta_0$ is assumed to be 0 in the analysis).}
\label{fig:sol_pcm}
\end{center}
\end{figure}

\begin{figure}[ht]
\begin{center}
\includegraphics[width=0.95\columnwidth]{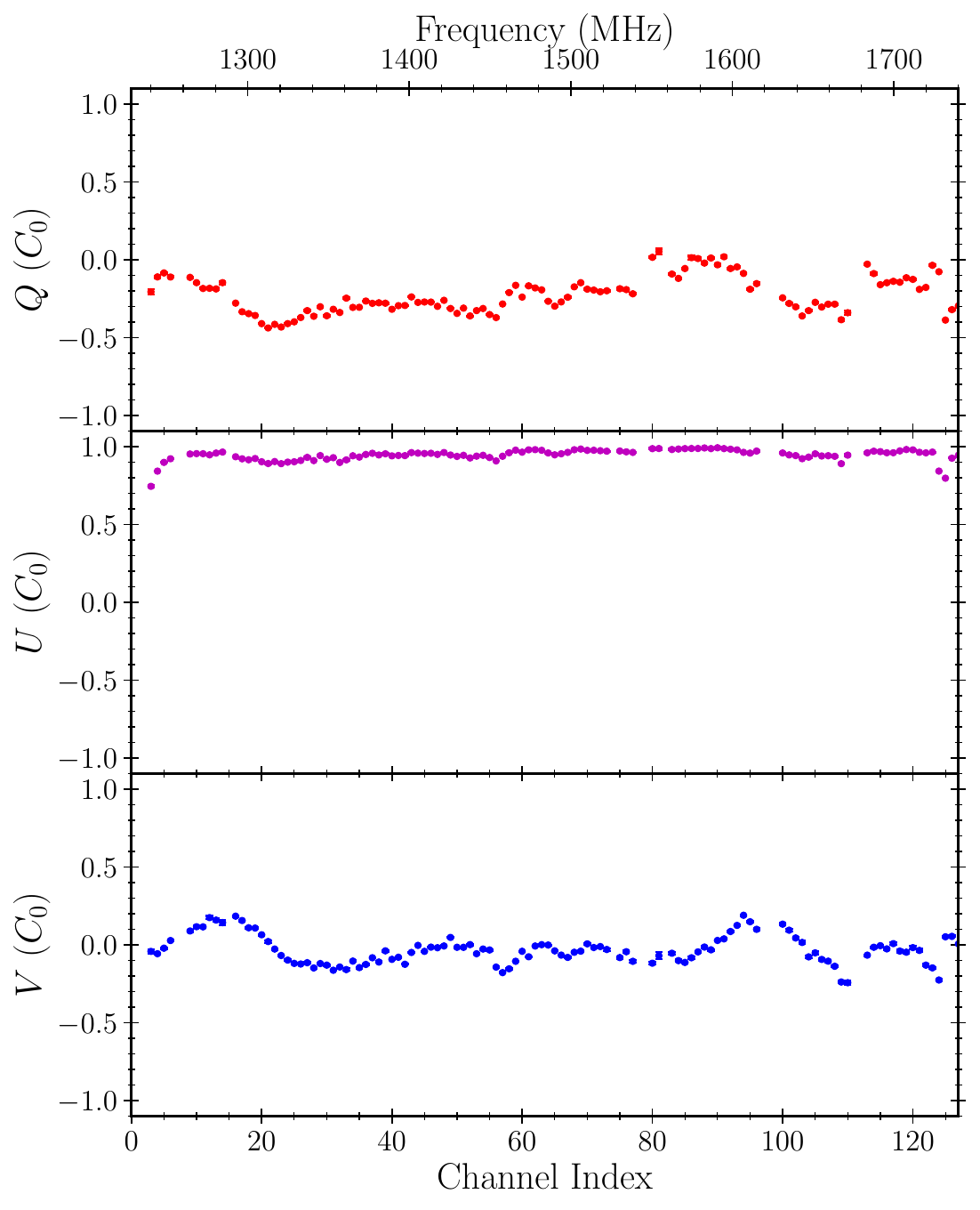}
\caption{$Q$, $U$, and $V$ Stokes parameters of the noise diode signal as a function of frequency, expressed in units of the flux density of the diode in the considered channel, $C_0$. The results plotted in this figure were determined by analyzing the noise diode observation shown in Figure~\ref{fig:cal_0742}, as part of the procedure for improving the calibration of NUPPI data described in Section~\ref{subsec:calib2}.}
\label{fig:cal_pcm}
\end{center}
\end{figure}

The gain, differential gain, and differential phase values, as well as the ellipticities and orientation parameters as determined from the \texttt{Reception} analysis, are plotted in Figure~\ref{fig:sol_pcm}. The Stokes parameters of the noise diode plotted as a function of channel index are shown in Figure~\ref{fig:cal_pcm}. The analysis finds significant non-orthogonality in most frequency channels, confirming the presence of cross-coupling between the polarization feeds. The gain and differential phase parameters are generally similar to those shown in Figure~\ref{fig:cal_0742}, determined from a simple \texttt{SingleAxis} analysis of the noise diode observation. Differential gains, however, strongly differ from those plotted in the latter figure in most channels. As can be seen from Figure~\ref{fig:cal_pcm}, the analysis confirms that the noise diode signal mainly consists of linear emission along the $U$ Stokes parameter, albeit with significant Stokes $Q$ and $V$ emission in most frequency channels, indicating that the noise diode signal is also not ideal. The ellipticities and orientations of the feeds, as well as the Stokes parameters of the noise diode plotted in Figures~\ref{fig:sol_pcm} and \ref{fig:cal_pcm}, need to be taken with a grain of salt since, as mentioned in the previous paragraph, strong assumptions about the non-orthogonality parameters needed to be made to constrain the mixing between $I$ and $V$, and between $Q$ and $U$. The analysis nevertheless demonstrates the existence of significant cross-coupling between the feeds and that the noise diode is most likely non-ideal.

The improved calibration should in principle lead to more accurate polarimetric results. We repeated the analysis of a selection of 1.4~GHz observations of MSPs with NUPPI presented in Section~\ref{subsec:calib1}, this time using improved calibration parameters obtained from the analysis of observations of PSR~J0742$-$2822 with feed horn rotation. For each of the MSP observations listed in Table~\ref{tab:obstable}, we used the PSRCHIVE command ``\texttt{pac}'' to calibrate the data using the second order calibration parameters from the closest-in-time observation of PSR~J0742$-$2822 and the noise diode observation conducted before the MSP observation. Equation~\ref{eq:calibration} was then used by \texttt{pac} to perform the calibration. The new polarimetric profiles are displayed in Figures~\ref{fig:pol_profs_1} to \ref{fig:pol_profs_4}, and the new PA measurements are shown in Figures~\ref{fig:PAs_1} and \ref{fig:PAs_2}. Additionally, in Table~\ref{tab:obstable} we list the profile S/N values as obtained when calibrating the data with the simple \texttt{SingleAxis} calibration scheme, and with the improved method. The polarimetric profiles formed with the improved calibration method are much closer to the \citet{Dai2015} profiles than those obtained with the generic calibration method, as can be seen from Figures~\ref{fig:pol_profs_1} to \ref{fig:pol_profs_4}. Similarly, the new PA values are very different from those obtained with the \texttt{SingleAxis} calibration method (see, for instance, the results for PSR~J1713+0747 in Figure~\ref{fig:PAs_1} or PSR~J1857+0943 in Figure~\ref{fig:PAs_2} for particularly discrepant results), and are also very close to the \citet{Dai2015} measurements. The latter results and those obtained with the improved calibration method are, in all cases highlighted in Figures~\ref{fig:PAs_1} and \ref{fig:PAs_2}, qualitatively consistent. The results are, however, not identical: some differences between the PA values are for example seen in the leading edge of the first peak of PSR~J1022+1001 or in the main peak of J1713+0747. It should be noted that the NUPPI observation of PSR~J1713+0747 was carried out about 60 days after the pulsar underwent an abrupt profile change on MJD 59321 \citep{Xu2021,Singha2021,Jennings2022}, while the reference data were taken several years before the event. As a consequence, the NUPPI and Parkes profiles are not expected to be identical. Residual calibration imperfections, or inaccurate RM values at the epochs of the NUPPI observations, could also produce discrepancies. However, the new NUPPI measurements generally agree well with the reference results, indicating that NUPPI polarimetric profiles calibrated with the improved scheme can be confidently used for studying the polarization emission properties of pulsars or determining RM values along their lines-of-sight. 

As can be seen from Table~\ref{tab:obstable}, the new S/N values are in almost all cases larger than those of the profiles calibrated with the \texttt{SingleAxis} method, with increases of about 20\% in the case of PSR~J1744$-$1134, or 30\% for PSR~J1024$-$0719. One notable exception appears to be that of PSR~J1022+1001, for which the new S/N value is significantly smaller with the improved calibration method, and for which the total intensity profile as obtained with the \texttt{SingleAxis} calibration method is closer to the reference profile than that obtained with the improved calibration. As mentioned earlier, we assumed that the mixing between Stokes $I$ and $V$ is zero when determining the polarimetric response. In PSR~J1022+1001, the maximum of the total intensity profile coincides with the maximum of Stokes $V$, in absolute value. We could thus expect significant differences in Stokes $I$ in parts of the profile where $|V|$ is large. It should also be noted that the pulse profile of PSR~J1022+1001 is known to exhibit significant time variations \citep[see e.g.,][]{Camilo1996,vanStraten2013,Padmanabh2021} over timescales of years, and could therefore differ in the NUPPI and Parkes observations considered here. Besides, the Stokes $Q$, $U$, and $V$ parameters as determined with the improved scheme are more consistent with the reference results. We therefore conclude that the new polarimetric profiles for PSR~J1022+1001 shown in Figure~\ref{fig:pol_profs_1} are a better representation of the true polarimetric properties of this pulsar, despite the lower S/N value. 

The new differential gain values shown in Figure~\ref{fig:sol_pcm} are significantly different from those derived from the analysis of the noise diode observation alone. S/N values are calculated using the total intensity profiles, which are formed by adding data from the two polarization feeds. Under the assumption of ideal feeds recording independent and identically distributed (i.i.d.) data, the noise levels in each feed have equal variance, and the S/N is proportional to the square root of the number of polarizations summed (two in this case). However, if data are calibrated using inaccurate differential gains and unaccounted cross-couplings then the i.i.d. assumption no longer applies, thereby reducing the effective number of polarizations added to less than two. We argue that the larger S/N values in Table~\ref{tab:obstable} are a consequence of the fact that the improved calibration, and particularly the substantial corrections to the differential gain parameter over large portions of the observed bandwidth, restored the i.i.d. assumption in most cases, resulting in larger effective sample sizes. \citet{Foster2015} studied the influence of inaccurate data calibration on pulsar observations and pulsar timing, based on simulations, finding that inaccurate calibration results in lower observed S/N than the ideal value. In their simulations, however, noise was added after the instrumental polarization transformation, whereas in our case the majority of the system noise was added before the differential gain and was thus subjected to the dominant boost transformation. Our conclusions are therefore consistent with and complementary to those of \citet{Foster2015}.

In general, the improved calibration method led to larger S/N values for the observations we analyzed. As a consequence, NUPPI TOAs extracted from these observations should have smaller uncertainties. Furthermore, regular NRT observations of PSR~J0742$-$2822 with horn rotation enable us to determine the polarimetric response of the instrument on a regular basis. Therefore, we also expect NUPPI TOAs to be more accurate, in addition to being more precise, as pulse profile distortions from inaccurate calibration should in principle be mitigated. Effects of the improved polarimetric calibration on NUPPI timing results are discussed in detail in the following section. 

%%%%%%%%%%%%%%%

\section{Pulsar timing with NUPPI}
\label{sec:timing}

\subsection{The 1.4~GHz NUPPI timing archive}
\label{subsec:data}

To assess the influence of the improved calibration on the quality of NUPPI timing data, we consider three of the MSPs analyzed in Section~\ref{sec:data}: PSRs~J1730$-$2304, J1744$-$1134, and J1857+0943, all three of them being well-timed PTA pulsars observed with high cadence since the beginning of NUPPI operations. The rationale for selecting these three MSPs is the following: of the MSPs considered in Section~\ref{sec:data}, PSR~J1744$-$1134 is the most polarized pulsar and should thus be particularly susceptible to calibration inaccuracies. It is also among the best targets used by PTAs to search for gravitational waves \citep[see e.g.,][]{Chen2021}. Conversely, PSR~J1857+0943 has the lowest degree of polarization in the sample according to \citet{Dai2015}, and thus represents an interesting test case. Finally, we chose to consider PSR~J1730$-$2304, for which \citet{vanStraten2013} predicted that the MTM technique (see Section~\ref{subsec:toas}) should determine TOAs with much ($\sim 30$\%) lower uncertainties than techniques that extract TOAs from total intensity profiles. In this subsection we present the 1.4~GHz NUPPI timing archive using PSR~J1744$-$1134 as an illustrative example, since this pulsar has been observed with high cadence with the NRT since NUPPI was commissioned, and with previous backends also.

\begin{figure*}[ht]
\begin{center}
\includegraphics[scale=0.67]{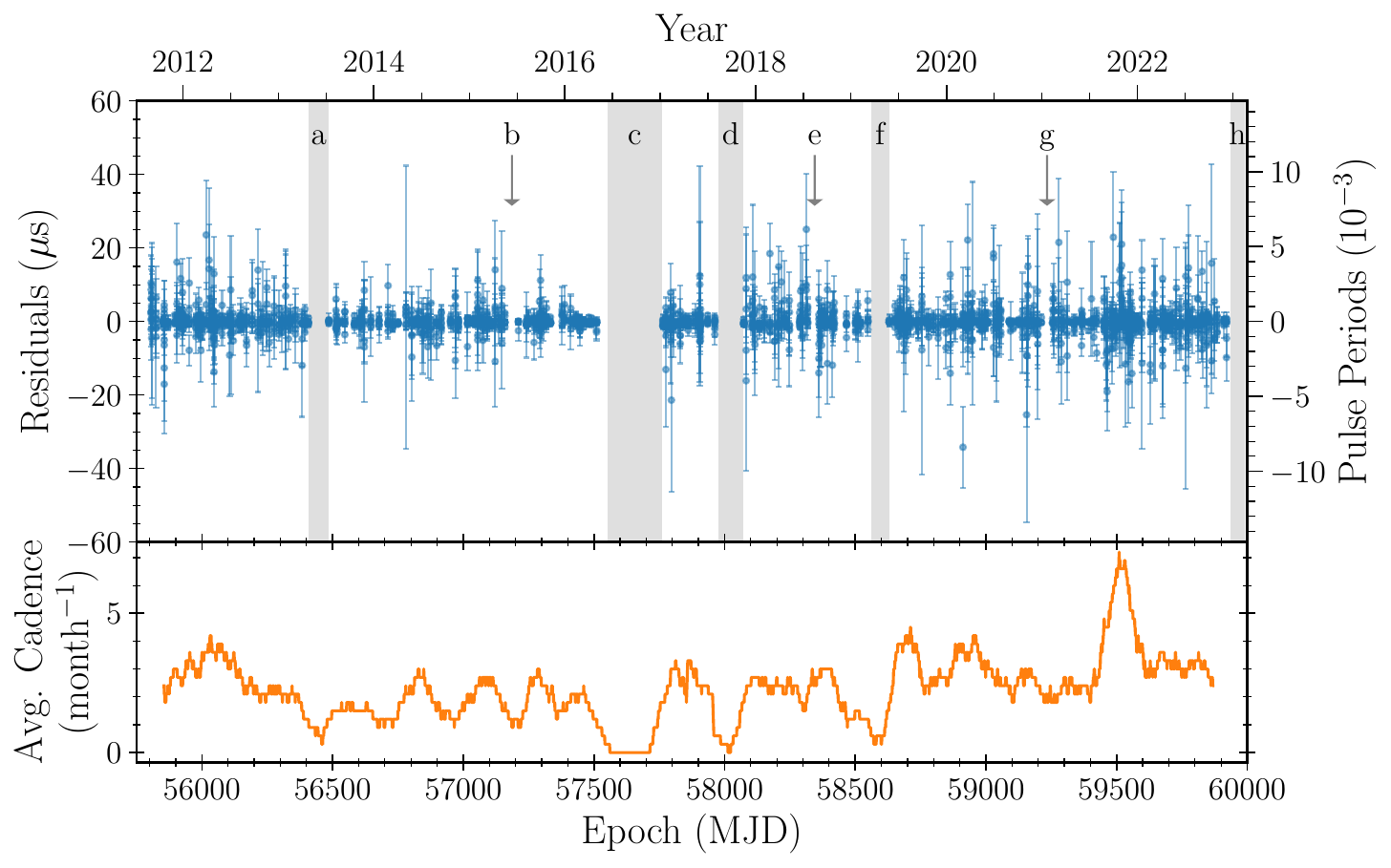}
\caption{NUPPI timing data on PSR~J1744$-$1134. \textit{(Top)} Timing residuals (\textit{i.e.,} differences between measured pulse arrival times and arrival times predicted by a model for the pulsar) as a function of time for the millisecond pulsar J1744$-$1134 observed with the Nan\c{c}ay Radio Telescope at the central frequency of 1.484~GHz using the NUPPI backend. The 512~MHz of frequency bandwidth recorded by NUPPI were split into eight sub-bands of 64~MHz each, and the observations were integrated in time. Pulse arrival times were extracted from the timing data using the MTM method, as further described in Section~\ref{sec:timing}. Time intervals when the NRT was inactive or when the low-frequency receiver was unavailable are highlighted with arrows (for interruptions shorter than 40 days) and shaded areas (when intervals were longer than 40 days). See Section~\ref{subsec:data} for details on the individual interruptions. \textit{(Bottom)} Average number of 1.4~GHz NUPPI observations of PSR~J1744$-$1134 per month as a function of time. The average cadence was estimated by counting the number of observations within a time window of 100-d centered on the considered date, and normalizing the obtained number to 30 days.}
\label{fig:1744_residuals}
\end{center}
\end{figure*}

In Figure~\ref{fig:1744_residuals} we show timing residuals for PSR~J1744$-$1134 as a function of time between August 2011 and December 2022, where the residuals correspond to the differences between the measured TOAs and the predictions from an accurate timing model for the pulsar. From the figure it can be seen that pulsar timing observations have been conducted at a relatively stable cadence between August 2011 and December 2022, with intervals of weeks to months when no timing observations were carried out. We here provide a list of the main interruptions, and what prevented observations from being carried out during those time intervals:

\begin{enumerate}[a)]
\item Painting activity in July 2013 and an intense observational campaign on the magnetar at a similar right ascension J1745$-$2900 \citep{Eatough2013} prevented J1744$-$1134 from being monitored normally between April and July 2013.

\item A fire in the electrical cabinet of the focal carriage and subsequent repairs prevented the NRT from observing between late May and early July 2015.

\item Due to the failure of an amplifier of the low-frequency receiver in June 2016, no 1.4~GHz observations could be carried out for several weeks. Incidentally, the failure of a pump in the hydraulic system of the focal carriage in September 2016 made it unable to move and track sources. The latter issue required heavy repairs, and normal observations resumed in January 2017.

\item Issues with the motion of the focal carriage along the track prevented NRT observations between August and November 2017.

\item The NRT was inactive for a few weeks in August 2018 due to electrical problems.

\item The failure of a pre-amplifier, which was replaced after a few weeks, prevented NRT observations with the low-frequency receiver from happening between late March and late May 2019.

\item A full repair of the focal track in early 2021 prevented the NRT from operating normally for a few weeks.

\item Electrical problems on the site caused the NRT to be inactive from late December 2022 to mid-February 2023.
\end{enumerate}

The above-mentioned interruptions are highlighted in the top panel of Figure~\ref{fig:1744_residuals}.

%%%%%%%%

\subsection{Timing data preparation}
\label{subsec:toas}

We selected NUPPI observations of PSRs~J1730$-$2304, J1744$-$1134, and J1857+0943 at 1.4 GHz, and cleaned the data of RFI using the \textsc{Surgical} method of the \texttt{CoastGuard} package. We then used the cleaned observations to prepare three distinct datasets for each pulsar. In the first dataset, the observations were left uncalibrated. In the second, observations were calibrated with the \texttt{SingleAxis} method described in Section~\ref{subsec:calib1}, \textit{i.e.,} using calibration parameters obtained from the analysis of the noise diode observations conducted prior to each of the pulsar observations, and assuming that the noise diode signal and the polarization feeds are ideal. In the third dataset, NUPPI observations were calibrated using the improved method described in Section~\ref{subsec:calib2}. In practice, the calibration parameters from the closest-in-time rotating horn observation of PSR~J0742$-$2822 were used, setting a maximum threshold for the time separation of 100 days. The first observation of J0742$-$2822 with feed horn rotation was conducted on MJD~58816; therefore, for NUPPI observations of PSRs~J1730$-$2304, J1744$-$1134, and J1857+0943 conducted before MJD~58716, a generic \texttt{SingleAxis} calibration was applied, as in the second dataset. Finally, all observations in the three datasets were integrated in time and frequency, forming eight frequency sub-bands of 64~MHz each.

\begin{figure*}[ht!]
\begin{center}
\includegraphics[width=0.95\columnwidth]{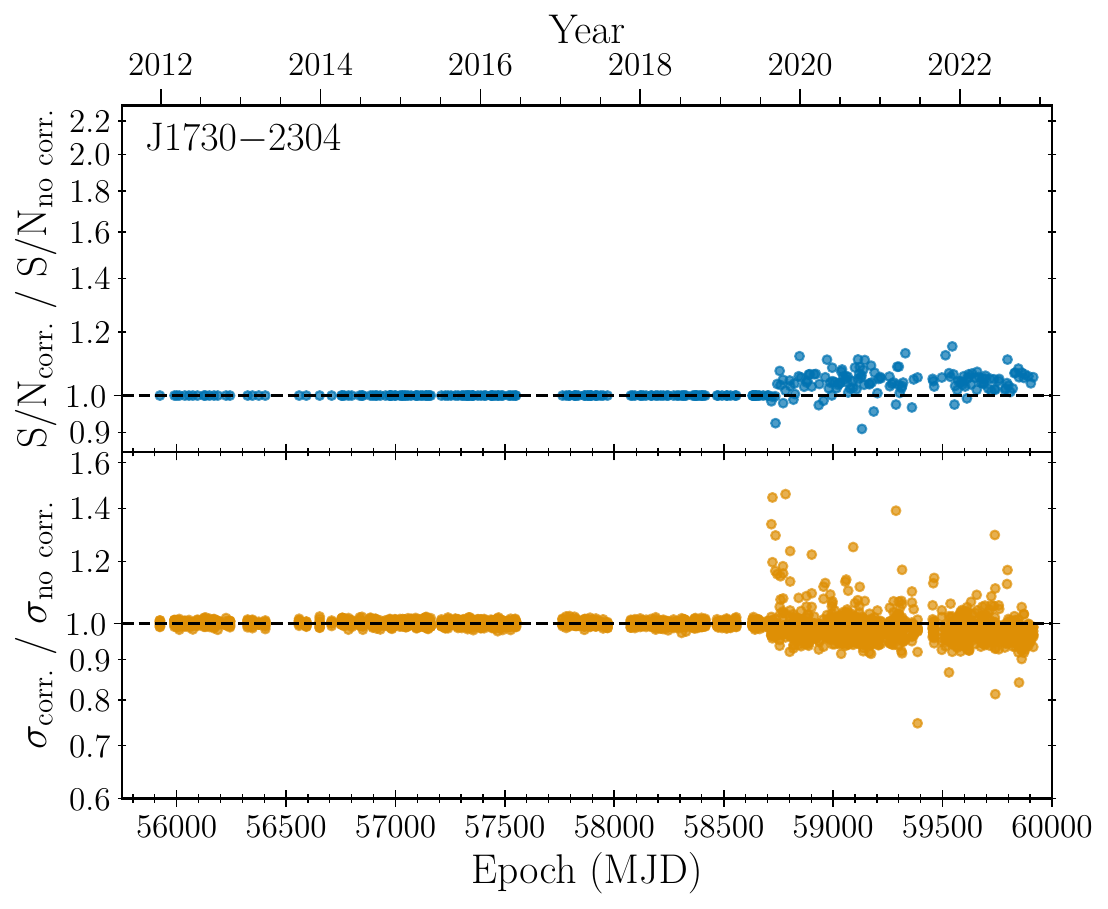}
\includegraphics[width=0.95\columnwidth]{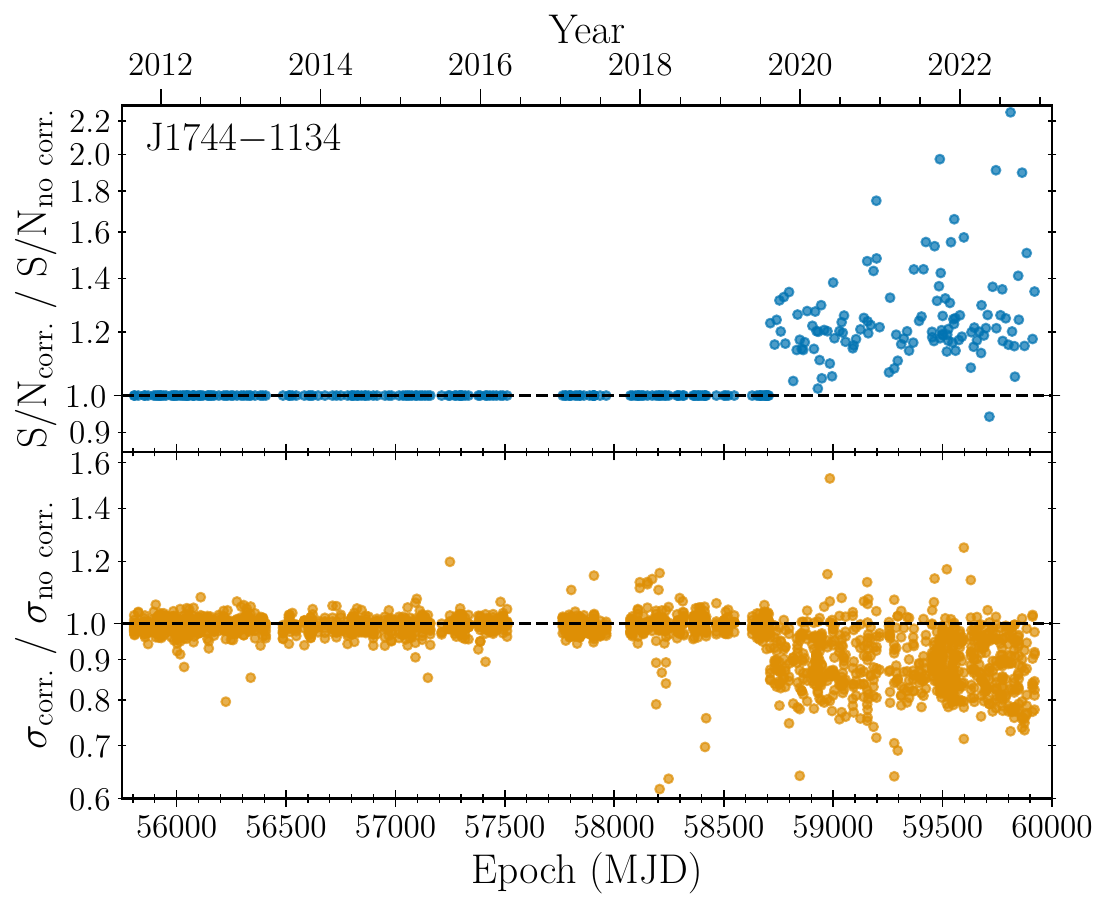}
\includegraphics[width=0.95\columnwidth]{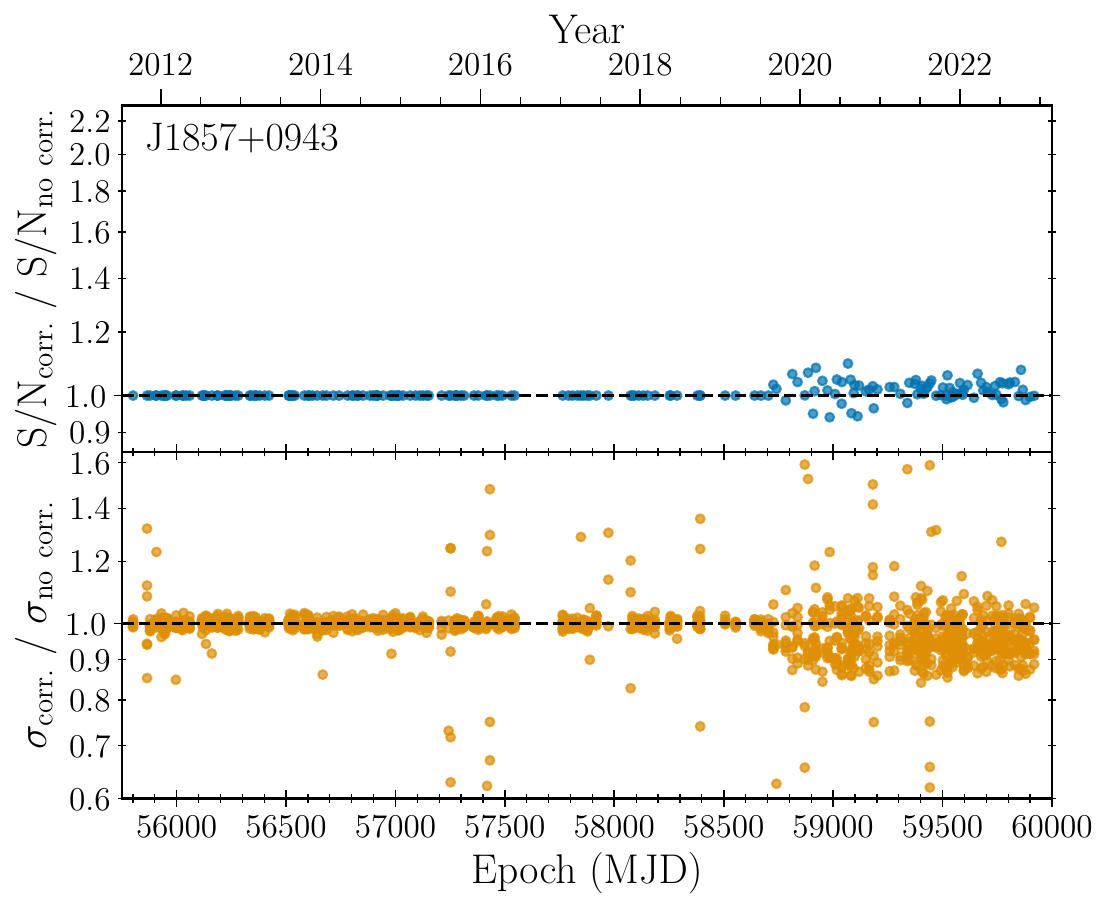}
\caption{Ratios of S/N and TOA uncertainty values for PSRs~J1730$-$2304, J1744$-$1134, and J1857+0943 as a function of time. The values are derived from NUPPI pulse profiles observed at 1.4~GHz and calibrated using the \texttt{SingleAxis} method (``no corr.'') and the improved method described in Section~\ref{subsec:calib2} (``corr.''). For each pulsar, the top panel (blue points) shows the ratios of the S/N values and the bottom panel (orange points) shows the ratios of the TOA uncertainties. Individual observations were fully integrated in time, and integrated in frequency to form eight sub-bands. TOAs were extracted from the observations using the MTM technique. See Section~\ref{subsec:toas} for additional details about the data preparation.}
\label{fig:snrs_uncs}
\end{center}
\end{figure*}

In Figure~\ref{fig:snrs_uncs}, we compare the S/N values of the total intensity pulse profiles from the second and third datasets. As expected, between MJDs 55800 and $\sim 58700$ the S/N values are identical, the calibration procedure being the same. However, it is clear from the figure that the improved calibration procedure generally leads to significantly larger S/N values as a consequence of accurate modeling of cross-couplings and polarization leakages, and particularly so for the highly polarized pulsar J1744$-$1134, with median increases of 4\%, 20\%, and 2\%, respectively, for PSRs~J1730$-$2304, J1744$-$1134, and J1857+0943.

In standard pulsar timing analyses, TOAs are extracted from the data by comparing total intensity profiles and a template profile for the pulsar at the considered frequency, using a cross-correlation algorithm. \citet{Wang2022} compared different methods for generating the template profiles, and found that for bright pulsars such as the ones considered here, templates formed by smoothing the addition of a selection of high S/N observations of the pulsar were preferred over other standard template preparation methods. \citet{Wang2022} further recommended the usage of the ``FDM'' method, as implemented in the PSRCHIVE tool ``\texttt{pat}'' for extracting TOAs from the observations. In the FDM method, the phase-gradient of the cross-power spectrum of the template profile and the observation are fitted in the Fourier domain \citep{Taylor1992}, as in the PGS method of \texttt{pat}, to determine the TOAs, but with TOA uncertainties estimated from a Monte Carlo Analysis. The FDM method was found by \citet{Wang2022} to produce TOAs and TOA uncertainty estimates that are generally more reliable than in the other tested methods. We adopted the recommendations of this systematic study of TOA creation procedures, and generated template profiles for each of the three datasets on the three pulsars by adding the total intensity profiles for the ten highest S/N observations made after MJD~58800 and then smoothing the summed profiles. The time selection was used to ensure that the template profile for the third dataset be formed by adding observations calibrated with the same procedure. TOAs for the three datasets were then extracted by comparing the observations with the respective template profiles, using the FDM method of \texttt{pat}.

As mentioned above, calibration observations of PSR~J0742$-$2822 have been conducted at Nan\c{c}ay since MJD~58816. We therefore do not have measurements of the full polarimetric response of the NRT before that date. In addition, as mentioned in Section~\ref{subsec:data}, various instrumental changes, such as the replacement of a pre-amplifier in 2019, occurred over the time span of the NUPPI dataset. Some of these instrumental changes likely modified the polarimetric response of the NRT, potentially leading to varying total intensity pulse profiles at different epochs. In addition to extracting TOAs from total intensity profiles using the FDM method, we thus also investigated the use of the MTM technique presented in \citet{vanStraten2006} and implemented in \texttt{pat}, to extract TOAs from the NUPPI data. Unlike the FDM method, the MTM technique uses all four Stokes parameters and models a transformation between the template profile and the considered observation, to reduce timing errors caused by instrumental distortions of the pulse profile. In this approach, observed polarization profiles represented by the coherency matrix, $\boldsymbol{\rho^\prime}$, are related to the template profile for the pulsar, $\boldsymbol{\rho}_0$, by the equation:

\begin{equation}
\boldsymbol{\rho^\prime} \left( \Phi \right) = \boldsymbol{J}\ \boldsymbol{\rho}_0 \left( \Phi - \Delta \Phi \right) \boldsymbol{J}^\dagger + \boldsymbol{\rho}_\mathrm{DC} + \boldsymbol{\rho}_\mathrm{noise} \left( \Phi \right),
\end{equation}

\noindent
where $\Phi$ is the pulse phase, $\Delta \Phi$ is the phase shift, $\boldsymbol{\rho}_\mathrm{DC}$ and $\boldsymbol{\rho}_\mathrm{noise}$ are the coherency matrices for the DC offset between the profiles and for the noise, and $\boldsymbol{J}$ is the polarimetric transformation between $\boldsymbol{\rho}_0$ and $\boldsymbol{\rho^\prime}$. By comparing the four Stokes parameters of the observations and of the template profile in the Fourier domain, the MTM method determines the $\Delta \Phi$ parameter and the Jones matrix, $\boldsymbol{J}$. We generated template profiles for each of the three datasets on the three MSPs by summing the ten highest S/N observations carried out after MJD~58800 and smoothing the added profiles, this time keeping all four Stokes parameters. The PSRCHIVE command \texttt{pat} was finally used to extract TOAs using these template profiles and the MTM algorithm. In the bottom panel of Figure~\ref{fig:snrs_uncs} we compare the uncertainties of TOAs extracted with the MTM method from the observations from the second and third datasets. Before MJD $\sim 58700$, the ratios are centered on one, but are not necessarily equal to one due to the fact that, although the procedure followed for calibrating the observations is the same for these epochs, the template profiles used for extracting the TOAs are different between the two datasets. After MJD $\sim 58700$, TOA uncertainties for the third dataset are generally smaller, as expected, given that the S/N values are generally higher with the improved calibration scheme.  

\begin{figure}[ht!]
\begin{center}
\includegraphics[width=0.95\columnwidth]{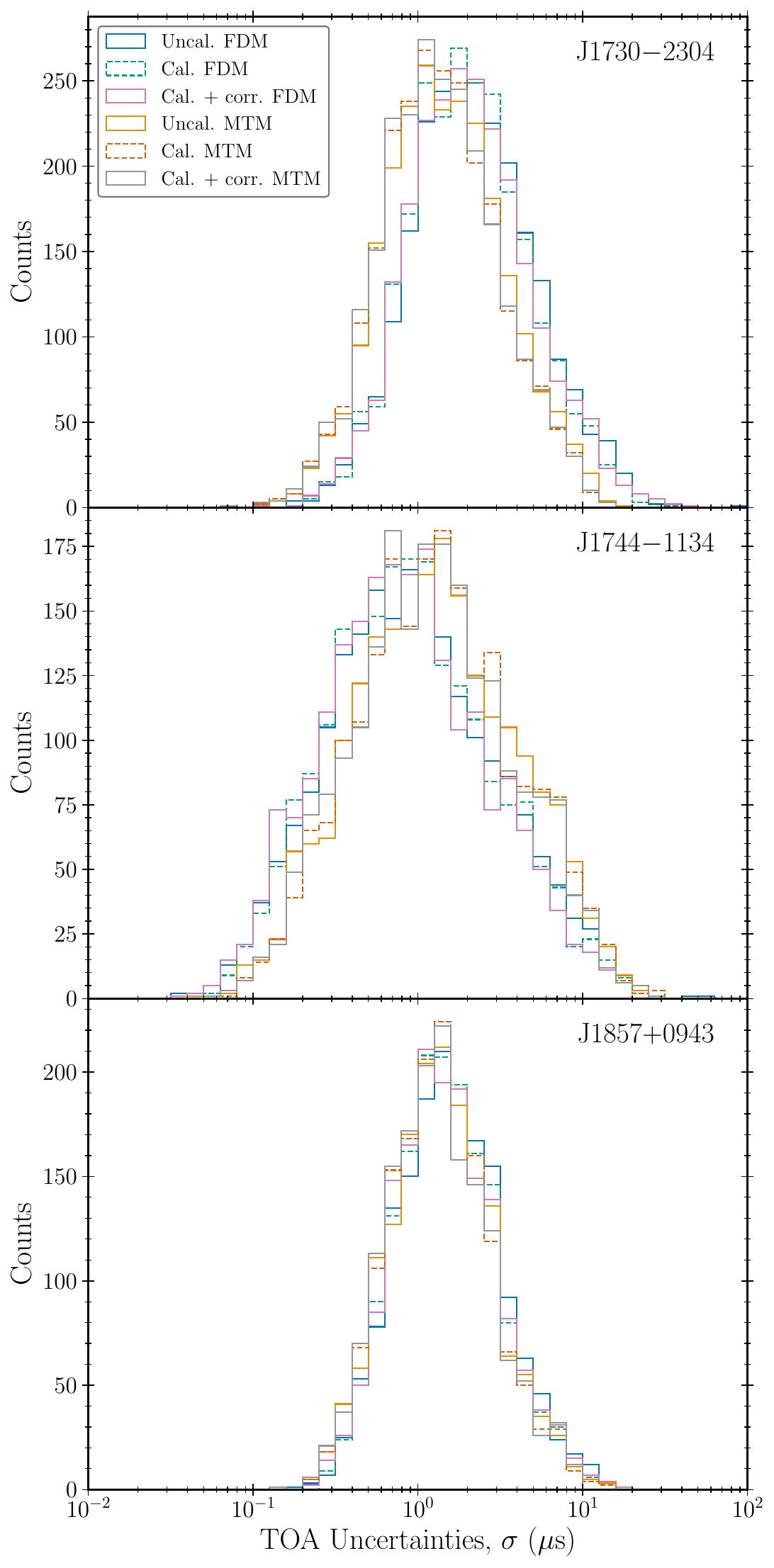}
\caption{Histograms of the TOA uncertainties, with TOAs extracted from uncalibrated NUPPI data (``Uncal.''), data calibrated using the \texttt{SingleAxis} method (``Cal.'') and data calibrated with the improved calibration method described in Section~\ref{subsec:calib2} (``Cal. + corr.''). Three of the TOA datasets were derived from the NUPPI data using the FDM algorithm; the other three were extracted using the MTM technique.}
\label{fig:hist_uncs}
\end{center}
\end{figure}

After the above-described TOA preparation steps, we obtained six different TOA datasets for each pulsar: TOAs extracted from uncalibrated NUPPI total intensity profiles with the FDM algorithm (we will refer to these TOA datasets as ``Uncal. FDM''), TOAs derived from uncalibrated NUPPI polarimetric profiles with the MTM algorithm (``Uncal. MTM''), TOAs obtained after a simple \texttt{SingleAxis} calibration with the FDM algorithm (``Cal. FDM'') and with the MTM method (``Cal. MTM''), and finally TOAs obtained after application of the improved calibration method with the FDM algorithm (``Cal. + corr. FDM'') and with the MTM method (``Cal. + corr. MTM''). Figure~\ref{fig:hist_uncs} presents histograms of the TOA uncertainties for the six different TOA datasets on each MSP. One can see that, as expected, the improved polarimetric calibration scheme leads to the smallest TOA uncertainties among the MTM and FDM datasets, while the largest TOA uncertainties are obtained for the uncalibrated data. Additionally, for the ``Cal. + corr.'' datasets, we find that the ratios of the median TOA uncertainties estimated by the MTM and FDM algorithms are respectively 0.68, 1.49, and 0.91 for PSRs~J1730$-$2304, J1744$-$1134, and J1857+0943. These values are very close to the theoretical expectations listed in Table~1 of \citet{vanStraten2013}, of 0.71, 1.56, and 0.89, respectively. For PSR~J1744$-$1134, which is almost 100\% linearly polarized \citep[see][]{Dai2015}, MTM finds larger TOA uncertainties due to large correlations between the phase shift and the parameters describing the polarization transformation. As can be seen from Table~1 of \citet{vanStraten2013}, MTM is expected to find smaller TOA uncertainties in most pulsars, because of their lower fractional polarizations.

%%%%%%%%

\subsection{Timing results}
\label{subsec:timing}

The ``Cal. + corr. MTM'' TOA datasets were built from better calibrated observations than the other datasets, and the TOAs were extracted using the MTM algorithm, which models putative profile distortions caused by calibration artifacts at the same time as it determines the TOAs. Of the different TOA datasets described in Section~\ref{subsec:toas}, they thus likely represent our best datasets for timing analyses of PSRs~J1730$-$2304, J1744$-$1134, and J1857+0943. We analyzed the three ``Cal. + corr. MTM'' datasets using the TEMPO2 pulsar timing package \citep{Hobbs2006}. Topocentric TOAs were converted to Barycentric Coordinate Time (TCB) using the DE440 solar system ephemeris described in \citet{Park2021}. The timing models were purposefully kept simple: for each pulsar we fitted for the sky coordinates, the proper motion and timing parallax, the pulsar's spin frequency and its first two time derivatives, the orbital parameters in the case of PSR~J1857+0943, and for the DM and its first two time derivatives. That is, we did not model any spin frequency or DM red noise components that could absorb timing instabilities caused by calibration issues. Similarly, we also did not model putative systematic errors in TOAs and/or in their uncertainties using EFAC and EQUAD parameters \citep[see e.g.,][]{Lentati2014}. The best-fit ephemerides are consistent with previously published results on these pulsars \citep[e.g.,][]{Desvignes2016}. In all cases the second spin frequency derivatives were significant, indicating the presence of spin noise in these pulsars.

\begin{figure*}[ht]
\begin{center}
\includegraphics[scale=0.5]{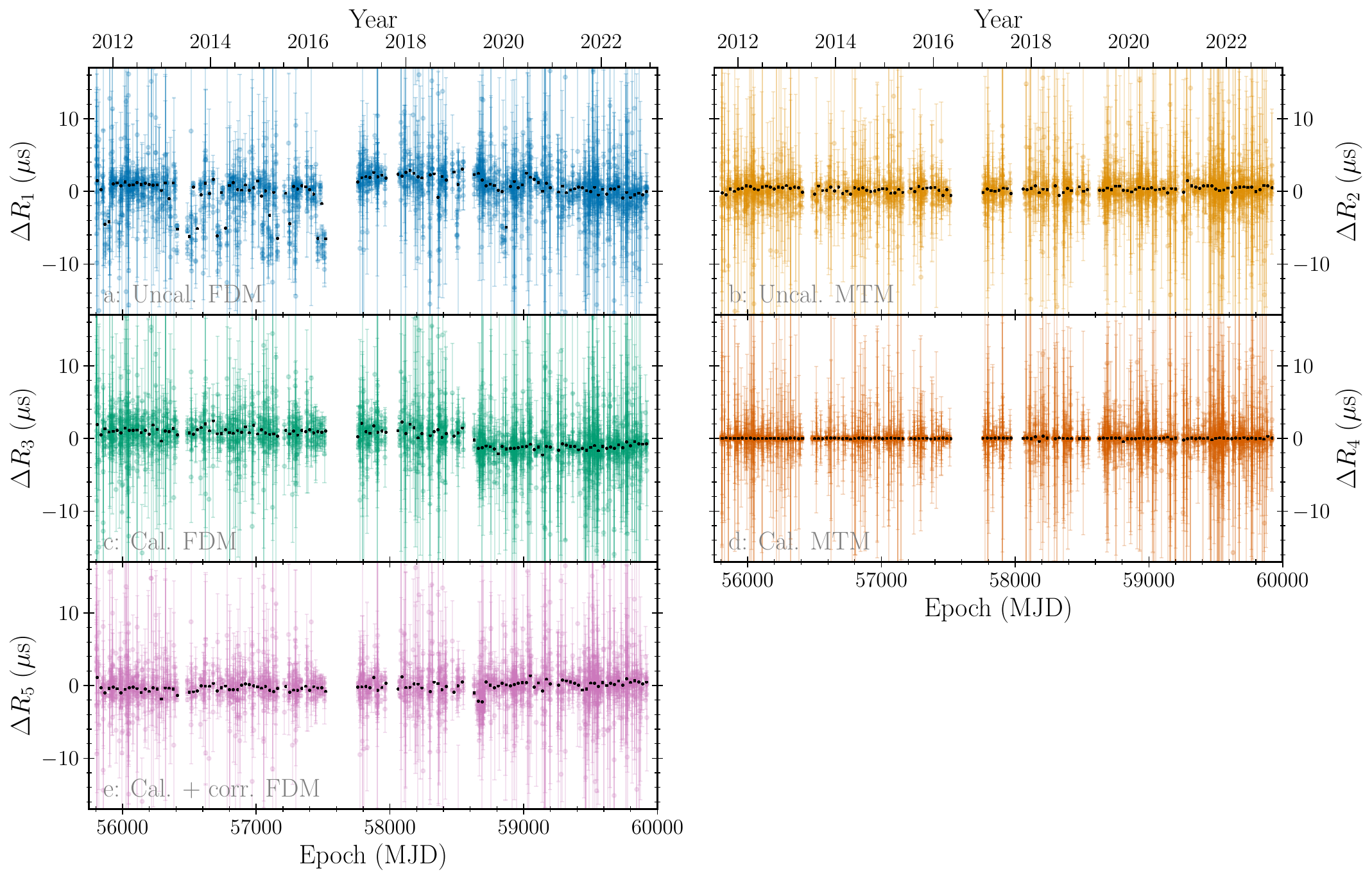}
\caption{Differences between timing residuals of PSR~J1744$-$1134 for the TOA datasets described in Section~\ref{subsec:toas}, and those for the ``Cal. + corr. MTM'' TOA dataset, as a function of time. In all cases, the residuals were determined using an ephemeris for the pulsar obtained by analyzing the ``Cal. + corr. MTM'' dataset with TEMPO2. Colored symbols represent the residual differences, and black dots show averages over 30-d sliding windows.}
\label{fig:residual_diffs}
\end{center}
\end{figure*}

The timing residuals for PSR~J1744$-$1134 as a function of time for the best-fit model and the ``Cal. + corr. MTM'' TOA dataset are plotted in Figure~\ref{fig:1744_residuals}. The weighted rms residual is $\sim 0.628$ $\mu$s, confirming PSR~J1744$-$1134 as one of the most precisely timed pulsars known currently. The reduced $\chi^2$ value is $\sim 2$, indicating that the timing data could be whitened further than what is done by our simple ephemeris for this pulsar. We used the best-fit model for each pulsar to determine timing residuals for the five other TOA datasets, and compared the timing residuals with those for the ``Cal. + corr. MTM'' dataset. In Figure~\ref{fig:residual_diffs} we show the differences in the timing residuals (\textit{i.e.,}, the differences between the residuals for the various TOA datasets, and those for the ``Cal. + corr. MTM'' dataset) for PSR~J1744$-$1134. As expected, the largest differences are seen for the ``Uncal. FDM'' dataset (see panel \textit{a}), with significant structures in the residual differences and many data points well away from zero. These differences reflect changes in the properties of the instrument that strongly influence the total intensity pulse profile of J1744$-$1134 at a given epoch, and can in particular make it inconsistent with the template profile used for extracting the TOAs. Interestingly, much less structure is seen for the ``Uncal. MTM'' TOAs (panel \textit{b}), demonstrating that the MTM algorithm manages to compensate for many of the profile changes induced by instrumental artifacts, albeit not completely, likely due to bandwidth depolarization \citep{vanStraten2002} in the 8 $\times$ 64~MHz frequency sub-bands. Much smaller residual differences are observed in the case of the ``Cal. FDM'' dataset (panel \textit{c}) than for the ``Uncal. FDM'' TOAs. However, a systematic offset is detected between the data taken before and after the pre-amplifier change at the NRT in 2019, showing that the total intensity pulse profiles are inconsistent in these two parts of the NUPPI archive. This offset is compensated for by the MTM algorithm, as can be seen from panel \textit{d} of Figure~\ref{fig:residual_diffs}. In the latter case, the residual differences are in all cases close to zero, indicating that for this pulsar the MTM algorithm finds similar results, independently of the calibration scheme. Finally, it can be seen from \textit{e} that the above-mentioned offset is not seen in data prepared with the improved polarimetric calibration scheme. However, as in panels \textit{a} and {c}, significant structure is seen in the residual differences, corresponding to cases where the MTM algorithm found differences between the observed pulse profiles and the template profile, which it compensated for. For the other two MSPs, J1730$-$1134 and J1857+0943, differences between timing residuals display the same behavior, albeit with more noise than in J1744$-$1134 due to the fact that they are less precisely timed objects.

\begin{figure*}[ht]
\begin{center}
\includegraphics[scale=0.5]{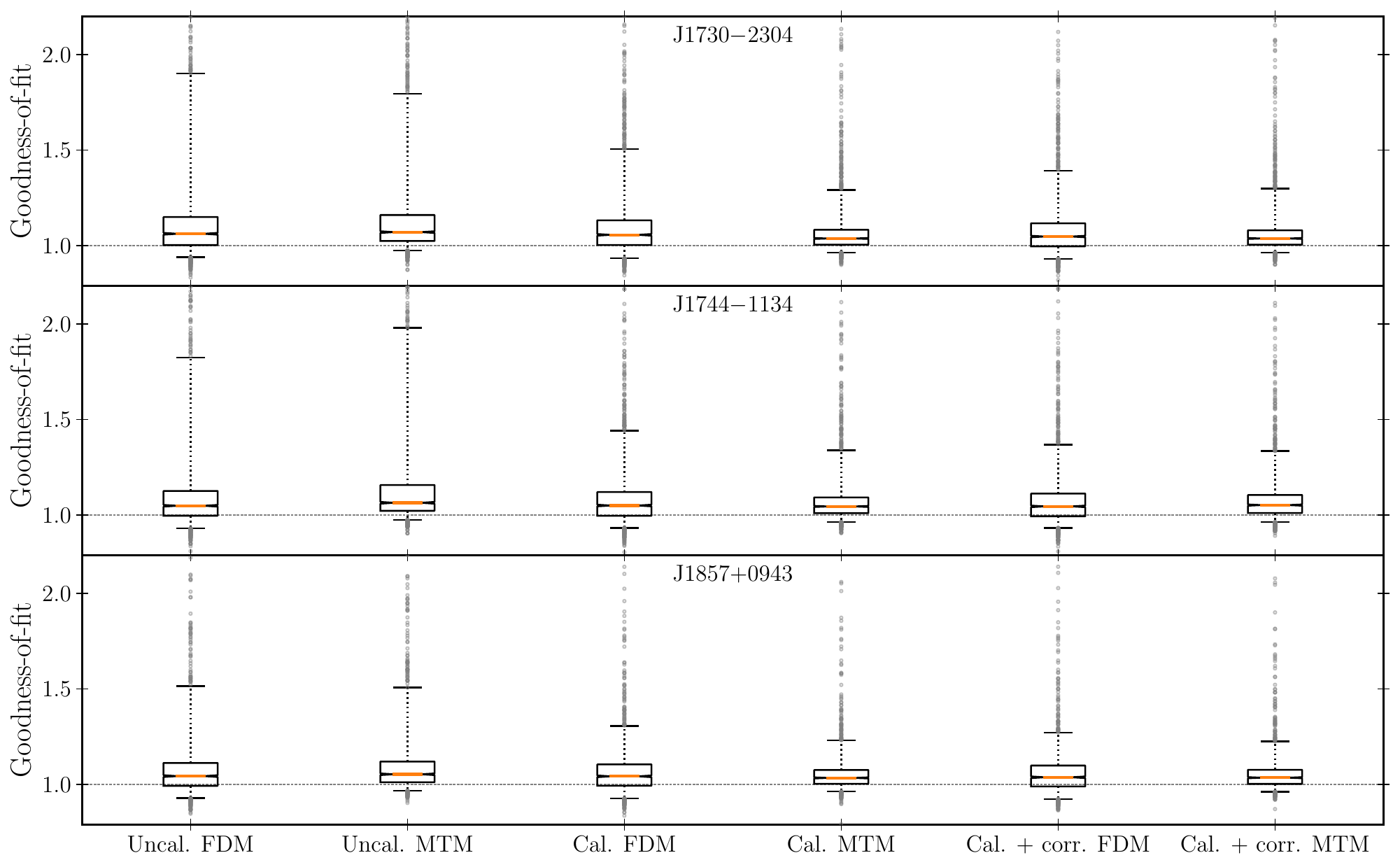}
\caption{Distributions of the reduced $\chi^2$ values of the residual profiles from template matching for each TOA, for the three pulsars and six dataset types considered in Section~\ref{subsec:timing}. Orange lines represent the median goodness-of-fit values, boxes encompass the first to third quartiles, whiskers encompass the 5$^\mathrm{th}$ and 95$^\mathrm{th}$ percentiles, and gray dots represent outliers.}
\label{fig:gofs}
\end{center}
\end{figure*}

Strictly speaking, although significant structure is seen in panel \textit{e}, we cannot tell from  Figure~\ref{fig:residual_diffs} if one of the two TOA datasets (``Cal. + corr. FDM'' or ``Cal. + corr. MTM'') is a better dataset than the other. We therefore conducted independent TEMPO2 analyses of all TOA datasets for the three pulsars to obtain best-fit timing models for each dataset and in turn determine weighted rms and reduced $\chi^2$ values. These parameters are listed in Table~\ref{tab:timing_res}. As expected, of the three TOA datasets extracted from total intensity pulse profiles with the FDM algorithm, the lowest weighted rms residual and reduced $\chi^2$ values are obtained for the dataset that is based on the best calibration method. The same conclusion also applies to TOA datasets derived using the MTM algorithm, although in this case, the ``Cal. MTM'' and ``Cal. + corr. MTM'' datasets have similar weighted rms residuals and reduced $\chi^2$ values. In Figure~\ref{fig:gofs} we show the distributions of the reduced $\chi^2$ values of the residual profiles from template matching for each pulsar and TOA dataset. The median goodness-of-fit values are in all cases close to unity, indicating that template matching fits are generally good. However, for all three pulsars, the ranges between the 5$^\mathrm{th}$ and 95$^\mathrm{th}$ percentiles are significantly larger for the uncalibrated data, and in the case of the calibrated data the ranges for MTM TOAs are systematically smaller than for FDM TOAs. Profile residuals therefore seem to provide a metric for assessing the correctness of the polarization calibration in this case. We note finally that in all three MSPs the weighted rms residuals and the reduced $\chi^2$ values from the timing analyses are higher for the ``Cal. + corr. FDM'' TOA datasets than for their MTM counterparts. This analysis confirms our initial assumption that the ``Cal. + corr. MTM'' datasets, which are based on better calibrated observations and include TOAs extracted with the MTM algorithm to account for potential changes of the instrumental response with time, are our best datasets for conducting the timing analysis of PSRs~J1730$-$2304, J1744$-$1134, and J1857+0943 with NUPPI. Additionally, the analysis independently confirms the conclusion from \citet{Foster2015}, based on simulations, that the MTM method produces better timing than methods that use total intensity profiles. \citet{vanStraten2006} noted that the MTM method may not perform well when the analyzed pulsar has a low degree of polarization or the estimated phase shift is highly covariant with the MTM model parameters that describe the unknown polarimetric transformation. In these situations where the FDM method or another method using the total intensity pulse profiles would be preferred over the MTM algorithm, our analysis has shown that better calibrated data lead to better timing datasets, as expected.

\begin{table*}
\caption[]{Weighted rms residual values and reduced $\chi^2$ values from individual timing analyses of the six TOA datasets described in Section~\ref{subsec:toas} with TEMPO2. The TOA datasets for PSRs~J1730$-$2304, J1744$-$1134, and J1857+0943, respectively, comprised 2379, 2085, and 1612 TOAs.}
\label{tab:timing_res}
\centering

\begin{tabular}{lcccccc}
\hline
\hline
TOA dataset & Uncal. FDM & Cal. FDM & Cal. + corr. FDM & Uncal. MTM & Cal. MTM & Cal. + corr. MTM \\
\hline
 & & & & & & \\
\multicolumn{7}{c}{J1730$-$2304}\vspace{0.2cm} \\
Weighted rms residual ($\mu$s) \dotfill & 2.213 & 1.710 & 1.604 & 1.090 & 0.974 & 0.960 \\
Reduced $\chi^2$ \dotfill & 4.091 & 2.426 & 2.259 & 2.149 & 1.839 & 1.843 \\
\hline
 & & & & & & \\
\multicolumn{7}{c}{J1744$-$1134}\vspace{0.2cm} \\
Weighted rms residual ($\mu$s) \dotfill & 1.612 & 1.042 & 0.705 & 0.905 & 0.635 & 0.628 \\
Reduced $\chi^2$ \dotfill & 27.951 & 10.034 & 5.648 & 4.480 & 1.895 & 2.041 \\
\hline
 & & & & & & \\
\multicolumn{7}{c}{J1857+0943}\vspace{0.2cm} \\
Weighted rms residual ($\mu$s) \dotfill & 1.364 & 1.359 & 1.280 & 1.272 & 1.166 & 1.120 \\
Reduced $\chi^2$ \dotfill & 2.001 & 1.958 & 1.903 & 2.203 & 1.898 & 1.817 \\
\hline
\end{tabular}
\end{table*}

%%%%%%%%%%%%%%%

\section{Summary and prospects}
\label{sec:summary}

Until recently, pulsar data recorded with the NUPPI backend of the NRT at 1.4~GHz were polarization-calibrated using the observations of a noise diode that injects a polarized reference signal into the feeds, assuming that both the instrumental response and the noise diode signal are ideal. In late 2019, we began observing the bright, highly linearly polarized pulsar J0742$-$2822 at a regular cadence, in a special observing mode in which the feed horn, which is fixed during normal pulsar observations, rotates by $\sim 180^\circ$ across the typically one-hour observation. Doing so enables us to more fully determine the polarimetric response of the telescope at the epochs of these observations as well as the actual Stokes parameters of the noise diode signal. We found significant cross-couplings between the polarization feeds and found that the noise diode does not emit 100\% linearly polarized emission along the Stokes $U$ parameter, as was assumed prior to this work. We applied these improved calibration parameters to the observations of a selection of MSPs with published polarimetric results, and found our new results to be consistent with those found by \citet{Dai2015} using the Parkes radio telescope. Finally, we analyzed NUPPI timing data on the MSPs 
J1730$-$2304, J1744$-$1134, and J1857+0943, and found that TOAs extracted from total intensity pulse profiles obtained with the improved calibration procedure have lower residual rms and reduced $\chi^2$ values than TOAs extracted from uncalibrated data or from data calibrated with the generic method used prior to this work. We also found that the MTM method for extracting TOAs from polarimetric profiles while compensating for potential calibration errors gives excellent results for these three MSPs, with TOA residuals that have lower rms and $\chi^2$ values than those obtained with TOAs extracted from total intensity profiles. It should be noted that, even though calibration errors are expected to affect the pulse profiles of pulsars differently depending on their polarization properties, they introduce time-correlated signals in the timing data of different pulsars, and thus contribute noise to searches for gravitational waves by PTAs. Correcting polarization calibration errors is thus critical to these searches.

The polarimetric response of the NRT at the epochs of the rotating horn observations and the properties of the reference source were determined by solving Equation~\ref{eq:rotation}. As explained in detail in \citet{vanStraten2004}, solutions to this equation are degenerate under commutation, and assumptions need to be made to constrain the mixing between $I$ and $V$ and between $Q$ and $U$. We assumed that the two feeds have equal ellipticities and that the misorientation of the first receptor is zero. A possibility for determining the individual receptor ellipticities would be to combine the rotating horn observations with observations of standard flux candles. In future work we will carry out such observations of standard candles, close in time to rotating horn observations, to determine the individual ellipticities across the bandwidth covered by NUPPI. Additionally, some of the properties of the NRT, such as the gain of the telescope, are known to vary with the declination of the observed source \citep[see e.g.,][]{Theureau2005}. In this work we have calibrated observations of MSPs at different sky locations using instrumental responses determined from observations of J0742$-$2822, which may or may not be optimal, given the potential dependency of some of the calibration parameters on declination or other pulsar parameters. This, and the possible variations of the instrumental response during observations (as noted in Section~\ref{subsec:calib2}), will also be investigated in future work. Pulsar observations conducted in a mode in which appropriate feed horn rotation would maintain a fixed parallactic angle across the observations may be informative in this regard. 

As mentioned in Section~\ref{sec:data}, the first pulsar observations with the NUPPI backend were conducted in August 2011, and regular observations of J0742$-$2822 with feed horn rotation began in late 2019. The improved calibration solutions derived from these later data cannot be directly applied to earlier observations because the response of the NRT likely evolved with the various instrumental changes that occurred since 2011. However, using the well-calibrated observations of pulsars observed since late 2019 as polarized reference sources, it is possible to model the instrumental response over the first eight years of NUPPI observations using Measurement Equation Template Matching \citep[METM;][]{vanStraten2013}. The improved calibration solutions for the NUPPI observation archive can be expected to lead to improved timing results, which is of great interest for pulsar timing studies and PTA analyses.

Observations of PSR~J0742$-$2822 with feed horn rotation are conducted at Nan\c{c}ay to overcome the fact that the NRT cannot sample wide ranges of parallactic angles when observing a given pulsar. The work presented in this article is thus relevant for any telescope affected by a similar limitation. The results obtained with the improved calibration procedure demonstrate that accurate polarization calibration is essential for any pulsar polarimetric or timing study, and the need for accurate calibration will get more and more acute for the increasingly sensitive radio telescopes of the future. 

%%%%%%%%%%%%%%%

\begin{acknowledgements}

We would like to thank the anonymous referee for very constructive and valuable comments that helped to improve the manuscript.

The Nan\c{c}ay Radio Observatory is operated by the Paris Observatory, associated with the French Centre National de la Recherche Scientifique (CNRS). We acknowledge financial support from the ``Programme National de Cosmologie et Galaxies'' (PNCG) and ``Programme National Hautes Energies'' (PNHE) of CNRS/INSU, France. 

Throughout this work, we made extensive use of the ATNF Pulsar Catalogue database \citep{Manchester2005}, available at: \url{https://www.atnf.csiro.au/research/pulsar/psrcat/}. 

Part of this research has also made use of the EPN Database of Pulsar Profiles maintained by the University of Manchester, available at: \url{http://www.jodrellbank.manchester.ac.uk/research/pulsar/Resources/epn/}. The published data displayed in Figures~\ref{fig:pol_profs_1} to \ref{fig:pol_profs_4} were obtained from the EPN database under the Creative Commons Attribution 4.0 International licence.

\end{acknowledgements}

%%%%%%%%%%%%%%%

\bibliographystyle{aa} 
\bibliography{nrt_polar_1.bib}

\end{document}